\documentclass[conference,10 pt]{IEEEtran}
\makeatletter
\def\ps@headings{%
\def\@oddhead{\mbox{}\scriptsize\rightmark \hfil \thepage}%
\def\@evenhead{\scriptsize\thepage \hfil\leftmark\mbox{}}%
\def\@oddfoot{}%
\def\@evenfoot{}}
\makeatother
\usepackage{times}

\usepackage{cite}
\usepackage{url}

   \usepackage[dvips]{graphicx}
   \DeclareGraphicsExtensions{.eps, .jpeg}

\usepackage[cmex10]{amsmath}

\ifCLASSINFOpdf
\else
\fi
   \usepackage[dvips]{graphicx}
   \DeclareGraphicsExtensions{.eps, .jpeg}

\usepackage{algorithmic}
\usepackage[tight,footnotesize]{subfigure}

\usepackage{pgf}

\newtheorem{mytheorem}{Theorem}

\newtheorem{corollary}{Corollary}
\newtheorem{lemma}{Lemma}

\newtheorem{definition}{Definition}

\newcommand{\graphsize}{0.38}
\newcommand{\biggergraphsize}{0.43}

\newcommand{\smallgraphsize}{0.29}  

\newcommand{\proofbox}{{\hfill\rule{2mm}{2mm} \vspace{\parskip} }} 

\makeatletter
\def\url@leostyle{%
  \@ifundefined{selectfont}{\def\UrlFont{\sf}}{\def\UrlFont{\small\ttfamily}}}
\makeatother
\usepackage{ifthen}
\newboolean{TechReport}
\setboolean{TechReport}{true}

\ifthenelse{\boolean{TechReport}}
{}
{}

\begin{document}
%
\ifthenelse{\boolean{TechReport}}
{\title{Establishing Multiple Survivable Connections (Extended Version)}} 
{\title{Establishing Multiple Survivable Connections}} 

\author{\IEEEauthorblockN{\large Michael Keslassy and Ariel Orda}
\IEEEauthorblockA{Department of Electrical Engineering\\
Technion - Israel Institute of Technology\\
Haifa 32000, Israel\\
\{keslassy@tx,ariel@ee\}.technion.ac.il }}


%


\maketitle

\begin{abstract}
We consider the establishment of connections in survivable networks,
where each connection is allocated with a working (primary) path and
a protection (backup) path. While many optimal solutions have been
found for the establishment of a single connection, only heuristics
have been proposed for the multi-connection case.

We investigate the algorithmic tractability of multi-connection
establishment problems, both in  online and offline frameworks.
First, we focus on the special case of two connections, and study
the tractability of three variants of the problem. We show that only
one variant is tractable, for which we establish an optimal solution
of polynomial complexity. For the other variants, which are shown to
be NP-hard, heuristic schemes are proposed. These heuristics are
designed based on the insight gained from our study. Through
simulations, we indicate the advantages of these heuristics over
more straightforward alternatives.

We extend our study to $K$ connections, for a fixed $K$ that is
larger than $2$. We prove the tractability of online and offline
variants of the problem, in which sharing links along backup paths
is allowed, and we establish corresponding  optimal solutions. On
the other hand, we show that, if the number of connections is not
fixed (i.e., $K$ is not $O(1)$), then the problem is, in general,
NP-hard.

\end{abstract}


%
\IEEEpeerreviewmaketitle

\section{Introduction}
\subsection{Motivation}

With the increasing use of telecommunication networks, the ability
of a network to recover from failures become a crucial requirement.
In particular, link failures can be catastrophic in networks that
are capable of transmitting very large amounts of data on each
link\cite{bhandari}. Accordingly, various routing schemes have been
proposed for supporting {\em network survivability}. Such schemes
handle link failures by rerouting connections through alternative
paths.

Survivable networks have been studied extensively in the context of
optical wavelength division multiplexing networks (WDM) and
multiprotocol label switching (MPLS) protection\cite{grover}. In WDM
networks, a link failure rarely happens and is in general repaired
before the next failure, hence justifying the employment of the {\em
single link failure model}. In this model, at most one link failure
happens at any given time. ``Working'' and ``protection'' lightpaths
are precomputed to ensure path restoration. In the MPLS context, a
``backup'' label switched path (LSP) is precomputed in case of a
path failure. The backup path allows active path restoration upon a
failure in the working (primary) path. Several schemes for the
selection of the working and backup LSPs have been
proposed\cite{mpls}.

Two main schemes in survivable networks are {\em 1+1 protection} and
{\em 1:1 protection}. In both schemes, the connection is composed of
two paths, namely the {\em primary path} and the {\em backup path}. In 1+1
protection, the destination node selects the best path at reception
of the data. In 1:1 protection, the backup path is used only if the
primary path that carries the data fails, allowing a backup sharing
between connections. For both schemes, the connection requests may
be known in advance (an {\em offline} framework) or not (an {\em
online} framework) \cite{wdm}.

The problem of establishing a survivable connection in a network is
vastly different when handling more than a single connection.
Indeed, optimal algorithms that are computationally efficient have
been established only for handling a single connection. Accordingly,
the purpose of this study is to explore the establishment of
multiple connections, and in particular investigate the possibility
(or lack thereof) of establishing optimal solutions that are
computationally tractable.

\subsection{Related Work}

As mentioned, the establishment of a single connection is
intrinsically more tractable in comparison to a multi-connection
establishment. Accordingly, much of the previous work on network
survivability focused on this simpler case. More specifically, most
studies focused on the establishment of a single connection that is
composed of two link-disjoint paths (see, e.g.,
\cite{bhandari,ho,li}). In contrast with this restriction, it has
been proposed in \cite{tunable} to establish connections with {\em
tunable survivability}, i.e., connections that can tolerate some
(nonzero) failure probability, hence allowing overlap between the
primary and backup paths. It has been shown in \cite{tunable} that
this added flexibility often results in a major improvement in the
usage of resources and quality of the solution. Several optimization
problems were proposed there, and corresponding optimal algorithms
of polynomial complexity were established, all under the single link
failure model and for the case of a single connection.

In case of multiple connections, previous studies focused on the
design of heuristic schemes (see \cite{grover} for a survey). An
important question when handling multiple connections is whether
there is some hierarchy among the connections, e.g. due to some
priority structure that forms part of a QoS architecture.

In particular, previous studies focused on the {\em Differentiated
Reliability} ({\em DiR}) model \cite{dir}. In this model,
connections belong to different classes that reflect their
importance. According to this classification, in case of a link
failure on its primary path, a higher-class connection is allowed to
preempt, through its backup path, links that are used by lower-class
connections. Thus, even with disjoint primary and backup paths and a
single failure in the network, some connections may fail, due to a
higher class connection preempting some of their dedicated
resources.

Adopting an online framework, some studies ordered the connections
according to the arrival time of the request: for each new request,
a restorable connection is sought such that previous connection
resources should be maintained. In particular, in \cite{drpi} a
partial information model was proposed for both contexts of
bandwidth guaranteed MPLS label switched paths and wavelength
switched paths (in networks with full wavelength conversion), and an
online heuristic, termed {\em Dynamic Routing with Partial
Information}, was presented. In \cite{mira}, the {\em Minimum
Interference Routing Algorithm} ({\em MIRA}) was proposed. More
specifically, MIRA is an online heuristic scheme that attempts to
allocate, to each incoming connection request, two paths that are
likely to minimize the interference with (the presently unknown)
future connection requests.

\subsection{Our Contribution}

We focus first on the basic, yet already complex case, of two
connections. After formulating the model, we specify several problem
variants, and proceed to analyze the computational complexity of
each. As a result, we establish an optimal solution of polynomial
complexity for one variant, while we prove the computational
intractability of the other variants. For the latter, we propose
heuristic solution schemes which, as indicated by simulations,
outperform more straightforward alternatives.

Next, we extend our analysis to the case of multiple (more than two)
connections. Assuming a fixed (i.e., $O(1)$) number of connections,
we establish an algorithm for online connection establishment that
allows to share resources over backup paths. To the best of our
knowledge, this is the first polynomial time algorithm that provides
an optimal solution to multi-connection establishment with sharing
of backup resources. In addition, we show that the offline version
of the problem is also computationally tractable. On the other hand,
we establish the intractability of several extensions of the
multi-connection problem, in particular the case of an unbounded
number of connections.

The remainder of the paper is organized as follows. Section II
presents our model and formulates the $K$ connection problem.
Section III focuses on two connections, i.e., $K=2$. Section IV
considers a larger ($K>2$) number of connections. Section V presents
a simulation study of our heuristic schemes. Finally, concluding
remarks are presented in section VI.

\ifthenelse{\boolean{TechReport}}
{}
{Due to space limits, some proofs and details are omitted and can be found in
\cite{tr}.}

\section{Model and Problem Formulation}

\subsection{The Model}

The network is represented by a graph $G=(V,E)$, where $V$ is the
set of nodes and $E$ the set of links. We denote by $n$ and $m$ the
number of nodes and links respectively, i.e. $n=|V|$ and $m=|E|$.
The links are assumed to be bidirectional. Thus, $G$ is undirected.
A (survivable) {\em connection} $c$ is characterized by a source
$s_c \in V$ and a destination $t_c \in V$. In order to cope with
link failures, a connection needs to be allocated with a {\em
primary path} and a {\em backup path}. The primary path of a
connection $c$ is the working path used to transfer its data from
the source $s_c$ to the destination $t_c$. The data is rerouted
through the backup path in case of a failure on the primary path
(1:1 protection scheme).

As explained, this study considers the establishment of some $K$
(survivable) connections, for a fixed (i.e., $O(1)$) value of $K$,
$K>1$. Particular attention is given to the case $K=2$.

We proceed to specify the failure model. Each link $e \in E$ of the
undirected graph $G$ is associated with a conditional failure
probability  $P_f (e)$, which is the probability that, given a
failure of a link in the network, $e$ is the failed link. Following
the single-link failure model, we assume that precisely one failure
occurs, i.e.:
\begin{eqnarray}
\sum_{e \in E} P_f (e) &=& \sum_{e \in E} P(\textrm{$e$ fails }
|\textrm{ exactly one link fails})\nonumber \\
&=& 1 \label{eq:A}\
\end{eqnarray}
For a path $p$ in $G$, we denote its (path) failure probability by
$P_f (p)$.  The path failure probability is, in fact, equal to the
sum of the failure probabilities of its links. Indeed, for a single
path $p$ in $G$:
\begin{eqnarray}
P_f (p) &=& \sum_{e \in E} P_f (e) \cdot P(\textrm{$p$ fails } | \textrm{ $e$ fails}) \nonumber \\
&=& \sum_{e \in p} P_f (e) \label{eq:A2}\
\end{eqnarray}
We assume that each connection $c$ is associated with a  {\em
Maximal Conditional Failure Probability}, $MCFP(c)$, which is the
maximal value of the failure probability that the connection can
sustain under the event of a single link failure. Accordingly, a new
connection request is approved only if it can be accommodated with
primary and backup paths that sustain its $MCFP$. For example, if
the $MCFP$ of the connection request is zero and all links have
nonzero failure probabilities, then the primary and backup paths
must be link-disjoint.

Denote by $P_f(c)$ the connection failure probability of a
connection $c$. A positive connection failure probability can be
caused due to two reasons: a link sharing between both paths of the
connection, or from a preemption by a connection of higher priority,
causing that part of its path allocation is diverted to the
preempting connection. We proceed to discuss these two cases. Following the
tunable survivability framework\cite{tunable}, a connection is
may share links between its primary and backup paths.
For a single connection $c$ in $G$, composed of a primary path $p_c$
and a backup path $b_c$, its connection failure probability is
expressed by:
\begin{eqnarray}
P_f (c) = P_f (p_c \cap b_c) 
= \sum_{e \in p_c, e \in b_c} P_f (e) \label{eq:B}\
\end{eqnarray}
We proceed to discuss connection preemption. Following the DiR
model\cite{dir}, we assume the $K$ connections are classified
according to a decreasing order of priority, such that $c_1$ is in
the class with the highest priority and $c_K$ in the class with the
lowest one. As per the DiR model, in case of failure on their
primary path, connections of higher priority may preempt, through
their backup path, links that form part of (any of the paths of)
connections with lower priority. Hence, such preemption is another
source of connection failure. Therefore, if a connection $c_i$ has
$MCFP(c_i)=0$, it should be accommodated with paths such that no
failure of a primary path of a connection with higher priority
(i.e., a connection $c_j$, for $j<i$) would result in the failure of
$c_i$.

\subsection{Problem Formulation}

We proceed to state the problem.

\begin{definition}
Given are: (i) a fixed (i.e. $O(1)$) value $K$, (ii) $K$
connections $c_1,..,c_K$, with source-destination pairs
$(s_1,t_1)$,..,$(s_K,t_K)$, respectively; (iii) maximal conditional
failure probabilities  $MCFP(c_1),..,MCFP(c_K)$ of $c_1,..,c_K$,
respectively. {\em The $K$ connection problem} (termed {\em Problem
KCP}) is the problem of finding primary and backup paths for the $K$
connections, $p_1$, $b_1$,.., $p_K$, $b_K$, such that the conditional failure
probabilities of the $K$ connections $c_1,..,c_K$ do not
exceed the respective maximal conditional failure probabilities
$MCFP(c_1),..,MCFP(c_K)$.
\end{definition}

{\em Problem 2CP} is the special case of Problem KCP for $K=2$.

\section{Two Connections}

In this section we focus on Problem 2CP. We begin  by
characterizing the problem through an optimization expression. Then,
we define and analyze the tractability of three variants of the
problem, and then specify and discuss corresponding algorithmic
solutions.

\subsection{Characterization of Problem 2CP}

We note that Problem 2CP is equivalent to the following problem:
given two connections $c_1$ and $c_2$ with their respective
source-destination pairs, and given the maximal conditional failure
probability of (only) the first connection, i.e., $MCFP(c_1)$ of
$c_1$, find primary and backup paths for the two connections such
that $MCFP(c_1)$ is observed and such that the failure probability
of the second connection is minimized. This problem can be
formulated by the following optimization expression:
\begin{eqnarray}
\min_{p_1,b_1,p_2,b_2} && P_f (p_2\cap b_2)+1_{b_1 \cap p_2 \neq \oslash,b_1 \cap b_2 \neq \oslash} \cdot P_f (p_1-b_1)\nonumber \\
&& + 1_{b_1 \cap b_2 = \oslash, b_1 \cap p_2 \neq \oslash, p_1 \cap b_2 \neq \oslash} \cdot P_f (p_1 \cap b_2) \nonumber \\
s.t. && P_f (c_1) \leq MFPC(c_1), p_1 \cap p_2 = \oslash
\label{eq:C}\
\end{eqnarray}
We proceed to
\ifthenelse{\boolean{TechReport}}
{explain the above expression.}
{ explain the above expression (a detailed
analysis can be found in \cite{tr}).}
Consider first the constraints. Clearly, the primary path $p_2$
cannot share a link with $p_1$. Thus, we can restrict ourselves to
link-disjoint primary paths. Consider now possible link sharing
between each of the following three pairs of paths: $b_1$ and $b_2$;
$b_1$ and $p_2$; or $p_1$ and $b_2$.
\ifthenelse{\boolean{TechReport}}
{In the Appendix we show that, among }
{In \cite{tr} we show that, among
}
the possible
eight cases (where in each case we either allow or disallow link
sharing between each of the above three pairs of paths), we can
restrict ourselves to the following three cases, all represented in
expression (\ref{eq:C}).
\begin{enumerate}
\ifthenelse{\boolean{TechReport}}
{\item
Shared backup case: $b_1$ and $b_2$ may intersect, while the
other two pairs are link-disjoint. We recall that the tunable
survivability framework is also adopted, hence intraconnection link
sharing (i.e., $p_1$ with $b_1$, or $p_2$ with $b_2$) is possible.
Thus, the failure probability of the second connection equals the
failure probability of that connection in the case that the first
connection does not exist. Indeed, in this case, the first
connection cannot preempt the second connection: if the link failure
occurs at $p_1$, then $b_1$ is employed but does not interfere with
$p_2$. If the link failure occurs at $p_2$, then $b_2$ is employed
and the first connection continues routing through $p_1$. The
failure probability of the second connection is in this case equal
to $P_f (p_2\cap b_2)$.
\item
Unavoidable first backup case: in this case, both paths $p_2$ and
$b_2$ of the second connection share links with the first backup
path $b_1$ but none with the first primary path $p_1$. The failure
probability of the second connection corresponds to the failure
probability in the case the first connection does not exist in
addition to the failure probability due to a potential preemption of
the first connection in case of a link failure on $p_1$ and not on
$b_1$. If the failure is on a link that is common to both paths
$p_1$ and $b_1$, then the first connection fails and the second
connection survives. Thus, preemption takes place only if the link
failure occurs on a link used only by $p_1$. The failure probability
of the second connection is in this case equal to $P_f (p_2\cap
b_2)+ P_f(p_1-b_1)$.
\item
Overlapped connection case: probably the less intuitive case, it is
when $b_2$ intersects with $p_1$ but not with $b_1$, while $p_2$
intersects with $b_1$. This situation can for example represent an
improvement to the second connection when otherwise an expensive
intersection with $b_1$ of both paths ($p_2$ and $b_2$) is
necessary. Thus, if $b_2$ intersects with $p_1$ but not with $b_1$,
any failure on links of $p_1 - b_2$ does not imply the failure of
the second connection anymore. The failure probability of the second
connection is in this case equal to $P_f (p_2)+ P_f (p_1 \cap b_2)$.
}
{\item Shared backup case: $b_1$ and $b_2$ may intersect,
while the other two pairs are link-disjoint. The failure
probability of the second connection is in this case
$P_f(c_2)=P_f (p_2\cap b_2)$.
\item
Unavoidable first backup case: in this case, both paths $p_2$ and
$b_2$ of the second connection share links with the first backup
path $b_1$ but none with the first primary path $p_1$. We get in
this case: $P_f(c_2)=P_f (p_2\cap b_2)+ P_f(p_1-b_1)$.
\item
Overlapped connection case: probably the less intuitive case, it is
when $b_2$ intersects with $p_1$ but not with $b_1$, while $p_2$
intersects with $b_1$. In this case, $P_f(c_2)=P_f (p_2)+ P_f
(p_1\cap b_2)$.}
\end{enumerate}

\subsection{Problem Variants}

We proceed to define three variants of Problem 2CP. These
variants differ according to their framework. We recall that in
an {\em online framework}, connections are established
successively (i.e., computed on the fly), whereas in an {\em
offline framework}, connections are established concurrently
(i.e., precomputed). In all variants, we require full
reliability for the first (higher priority) connection.

In the first problem variant, termed Problem 2CP-1, we work under an
online framework, i.e., the first connection has been established
and we need to allocate paths to the second connection. Problem
2CP-2 considers the same framework as 2CP-1 but adds flexibility to
the solution, in the sense that the backup path of the first
connection, $b_1$, can be rerouted upon arrival of the second
connection. Lastly, Problem 2CP-3 describes an offline framework,
where both connections need to be established concurrently.

\begin{definition}
Given the primary and backup paths of the first connection $c_1$,
namely $p_1$ and $b_1$, correspondingly, and given that
$MCFP(c_1)=0$, {\em Problem 2CP-1} is the problem of finding paths
for the second connection that minimize its failure probability.
\end{definition}
As we shall show, the problem can be solved by an efficient,
polynomial time algorithm. The minimization of the failure
probability of the second connection can be expressed as:
\begin{eqnarray}
\min_{p_2,b_2} && P_f (p_2\cap b_2)+1_{b_1 \cap p_2 \neq \oslash,b_1 \cap b_2 \neq \oslash} \cdot P_f (p_1-b_1) \nonumber \\
&& + 1_{b_1 \cap b_2 = \oslash, b_1 \cap p_2 \neq \oslash, p_1 \cap b_2 \neq \oslash} \cdot P_f (p_1 \cap b_2) \nonumber \\
s.t. && p_1 \cap p_2 = \oslash \label{eq:C1}\
\end{eqnarray}
\begin{definition}
Given the primary path $p_1$ of the first connection $c_1$, and
given that $MCFP(c_1)=0$, {\em Problem 2CP-2} is the problem of
finding a backup path $b_1$ for the first connection and primary and
backup paths $p_2$ and $b_2$, correspondingly, for the second
connection, that minimize the failure probability of the second
connection.
\end{definition}
We shall establish the computational intractability of the problem.
The minimization of the connection failure probability of the second
connection is now:
\begin{eqnarray}
\min_{b_1,p_2,b_2} && P_f (p_2\cap b_2)+1_{b_1 \cap p_2 \neq \oslash,b_1 \cap b_2 \neq \oslash} \cdot P_f (p_1-b_1) \nonumber \\
&& + 1_{b_1 \cap b_2 = \oslash, b_1 \cap p_2 \neq \oslash, p_1 \cap b_2 \neq \oslash} \cdot P_f (p_1 \cap b_2)\nonumber \\
s.t. && P_f (c_1) \leq 0, p_1 \cap p_2 = \oslash \label{eq:C3}\
\end{eqnarray}
\begin{definition}
{\em Problem 2CP-3} is the problem of finding primary and backup
paths both for the first and for the second connections, $c_1$ and
$c_2$, such that the first connection attains full reliability
(i.e., $MCFP(c_1)=0$) and such that the failure probability of the
second connection is minimized.
\end{definition}
Again, we shall establish that the problem is computationally
intractable. The respective minimization expression is:
\begin{eqnarray}
\min_{p_1,b_1,p_2,b_2} && P_f (p_2\cap b_2)+1_{b_1 \cap p_2 \neq \oslash,b_1 \cap b_2 \neq \oslash} \cdot P_f (p_1-b_1) \nonumber \\
&& + 1_{b_1 \cap b_2 = \oslash, b_1 \cap p_2 \neq \oslash, p_1 \cap b_2 \neq \oslash} \cdot P_f (p_1 \cap b_2) \nonumber \\
s.t. && P_f (c_1) \leq 0, p_1 \cap p_2 = \oslash \label{eq:C4}\
\end{eqnarray}

\subsection{Complexity Analysis of the Three Variants}
We proceed to determine the tractability of the three variants of
Problem 2CP. In addition, we specify and validate a computationally
efficient algorithm that solves Problem 2CP-1, and present heuristic
solutions for (the intractable) Problems 2CP-2 and 2CP-3.

Before proceeding, we
\ifthenelse{\boolean{TechReport}}
{present a useful decomposition and}
{}
begin with the following definition. In a connected graph $G$, a
{\em bridge link} is a link whose removal generates a disconnected
graph. We say that the bridge link {\em separates} a node $u$ from a
node $v$ if, after its removal, there is no path from $u$ to $v$.

\ifthenelse{\boolean{TechReport}}
{We describe now a process of decomposition of a single-link (i.e.,
single-edge) connected graph into components connected by
bridge links separating two specified nodes.

\begin{definition}
Given an undirected single-link connected graph $G=(V,E)$, and two
nodes $s$ and $t$ in $V$, {\em Decomposition D1} is the
decomposition of $G$ into $r$ ($r\geq 1$) successive connected
components, interlinked by the bridge links of $E$ that separate $s$
from $t$. We successively denote these $r$ components
$C_1,..,C_{r+1}$ such that $s$ is in $C_1$ and $t$ is in $C_{r+1}$.
\end{definition}

Figure~\ref{fig:R1} depicts a graph decomposition into four
connected components $C_1, C_2, C_3$ and $C_4$, and three bridge
links $e_1, e_2$ and $e_3$. If $G$ is disconnected and $s$ and $t$
are in the same connected component $C$, we can apply this
decomposition directly to $C$.
\begin{figure}
\centering
\includegraphics[width=\smallgraphsize  \textwidth]{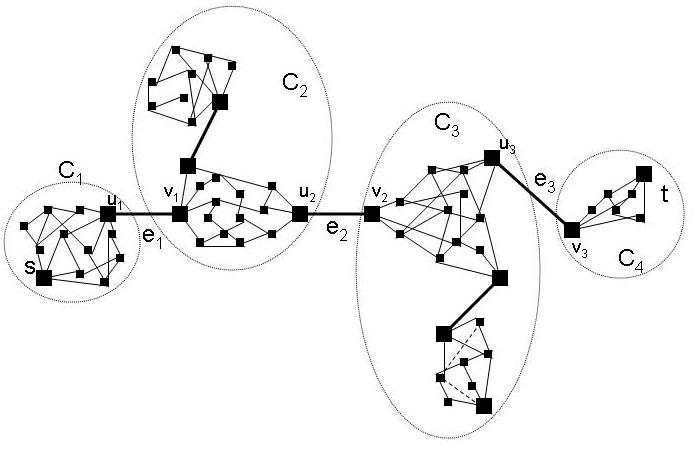}
\caption{Decomposition D1.}
\label{fig:R1}
\end{figure}

With the above decomposition scheme at hand, we }
{\begin{figure}
\centering
\includegraphics[width=\biggergraphsize  \textwidth]{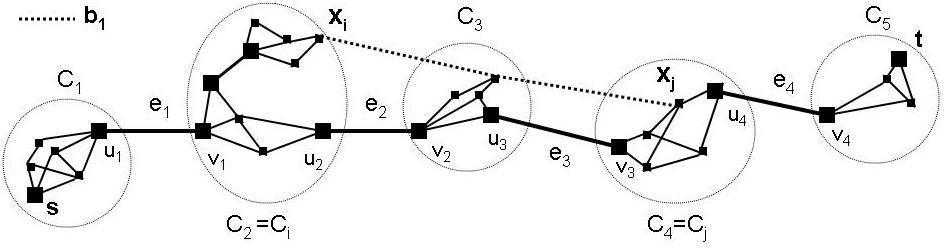}
\caption{The decomposition in the second step of Algorithm SCA.}
\label{fig:R1}
\end{figure}
First we
}
present an algorithm, termed Survivable Connection Addition (SCA),
that returns an optimal solution to Problem 2CP-1 within polynomial
time.

{\bf\em Algorithm SCA:}

\textbf{First step.} We consider an instance $(G,s_1,t_1,s_2,t_2)$
with the same notations as above, such that the links of $G$ have a
weight denoting their failure probability. In the graph $G_2$
representing the original graph $G$ without the links of $p_1$ and
$b_1$, enumerate the bridge links, $e_1,..,e_r$, that separate $s_2$
from $t_2$; these are the common links between the primary and the
backup paths obtained by the tunable survivability algorithm of
\cite{tunable}. If $s_2$ and $t_2$ are not in the same connected
component of $G_2$, go directly to the fourth step.

\textbf{Second step.} Decompose the joint connected component of
$s_2$ and $t_2$ in $G_2$ into components $C_1,..,C_{r+1}$ separated
by the bridge links $e_1,..,e_r$.
\ifthenelse{\boolean{TechReport}}
{- (i.e. Decomposition D1).}
{}
For $1\leq l\leq r$, each bridge link $e_l$ has a pair of endpoints
$(v_l, u_{l+1})$ such that $v_l$ is in $C_l$ and $u_{l+1}$ in
$C_{l+1}$. Define $u_1$ as $s_2$ and $v_{l+1}$ as $t_2$. Denote by
$C_i$ (respectively, $C_j$) the closest component with a common node
with $b_1$ from $C_1$ (respectively, $C_{r+1}$), if there are such
components, and denote by $x_i$ (respect., $x_j$) a node of $b_1$ in
$C_i$ (respect., $C_j$); see an example in Figure~\ref{fig:R1}.

\textbf{Third step.} Consider the primary and backup paths obtained
by \cite{tunable} in $G_2$ with $s_2$ and $t_2$ as source and
destination nodes. If $x_i$ and $x_j$ are in the same component, the
algorithm returns this solution. Otherwise, the solution is returned
at the end of this step. Add a fictitious link between these two
nodes and compute two link-disjoint paths $p_a$ and $p_b$ from $u_i$
to $v_j$ in the new graph. Without loss of generality, we may assume
that $p_a$ does not use the fictitious link. In particular, $p_a$
replaces the partial path of the primary path obtained by
\cite{tunable} between $u_i$ and $v_j$. The partial path of the
backup path obtained by \cite{tunable} between $u_i$ and $v_j$ is
also replaced with three different partial paths. Between $u_i$ and
$x_i$ the partial path is replaced by the partial path of $p_b$
between these two nodes. Between $x_i$ and $x_j$, the links of the
first backup path $b_1$ are used; between $x_j$ and $v_j$ the
partial path of $p_b$ between these two nodes is used.

\textbf{Fourth step.} The graph $G_1$ represents the original
graph $G$ without the links of $p_1$. If $s_2$ and $t_2$ are
not in the same connected component of $G_1$, no feasible
solution can be found. We denote by $q_1$ the value of the
failure probability of the second connection obtained by
\cite{tunable} for a connection request from $s_2$ to $t_2$ in
$G_1$.

\textbf{Fifth step.} The primary path $p_2$ is obtained as the
shortest path from $s_2$ to $t_2$ in $G_1$, where each link has the
same weight as in $G$. The graph $G'_1=(V'_1,E'_1)$ represents the
original graph $G$ without the links of $b_1$, such that every link
in $E'_1$ has a zero weight except the links of $p_1$, which keep
their original weights. The backup path $b_1$ is the shortest path
from $s_2$ to $t_2$ in $G'_1$. We denote by $q_2$ the failure
probability of the second connection composed of these two paths.

\textbf{Sixth step.} If $q_2 < q_1$, return the paths $p_2$ and
$b_2$. Otherwise, return the paths obtained by \cite{tunable} for a
connection request from $s_2$ to $t_2$ in $G_1$.

\begin{mytheorem}
\label{thm: Th A} Algorithm SCA solves Problem 2CP-1 in
$O(n~\cdot~m)$ steps.
\end{mytheorem}

\textbf{Proof:}
\ifthenelse{\boolean{TechReport}}
{We consider the three generic cases (shared backup, unavoidable
first backup, overlapped connections) that corresponds to the
optimization expression (\ref{eq:C1}), as described above, and we
analyze the optimality of the solution obtained by Algorithm SCA
under each case, as well as the corresponding computational
complexity.

Consider first the shared backup case: $p_1$ has no superposition
with other paths, while $b_1$ potentially has  some superposition
with only $b_2$. We prove here that if we can find two paths for the
second connection respecting the overlapping constraints of the
shared backup case, this solution globally minimizes the failure
probability of the second connection amongst all possible solutions.

Since there is no superposition with $p_1$, we can delete its links
from the graph to get a new graph $G_1$. In a one-link connected
graph, we can compute the minimal failure probability of a
connection as the sum of the bridge links between the source and the
destination as indicated by expression (3). We obtained a graph
called $G_2$ by deleting the links of $b_1$. If $s_2$ and $t_2$ are
not in the same connected component in $G_2$, no solution to the
shared backup case can be found, since the second primary path $p_2$
should be link disjoint with both of the paths of the first
connection. Thus the optimal solution can be found only within the
two other cases (first step of SCA).

Otherwise, if $s_2$ and $t_2$ share the same connected component in
$G_2$, we proceed with decomposition D1, listing the bridge links
$\{e_1,..,e_r \}$ of $G_2$ that separate $s_2$ from $t_2$ in
successive connected components $\{C_1,..,C_{r+1} \}$, as depicted
in Figure~\ref{fig:R1}. $s_2$ is located in $C_1$, and $t_2$ in
$C_{r+1}$. We denote by $C_i$ and $C_j$ the components of lowest and
highest index that have a common node ($x_i$ and $x_j$) with $b_1$
(Second step of SCA).

We claim that we can find two link-disjoint paths from $s_2$ to
$t_2$ that form a connection respecting the shared backup case
constraints and such that its failure probability equals the
sum of the failure probabilities of the bridge links that are
not between $C_i$ and $C_j$ (i.e. $\sum_{l=1}^{i-1} P_f (e_l) +
\sum_{l=j+1}^{r} P_f (e_l)$). Indeed, if we add a fictitious
link between $x_i$ and $x_j$, which stands for the partial path
of $b_1$ that  is forbidden to $p_2$ in this configuration,
then in the new graph the bridge links that separate $s_2$ from
$t_2$ are the same except the ones between $C_i$ and $C_j$.
Thus, in this new graph, we select two paths that minimize the
connection failure probability. Only one of them uses the
additional link, and it should be $b_2$ (third step of SCA). In
addition, this failure probability is minimal amongst the
different cases: in the unavoidable first backup case, the same
links have to be shared by the paths of the second connection
since they are also bridge links in $G_1$. In the overlapped
connection case, the failure probability of the second
connection includes the one of the primary path $p_2$ and by
its construction, $p_2$ uses all the bridge links between $s_2$
and $t_2$ in $G_1$. As a consequence, its failure probability
is superior to that obtained by the shared backup case.

We consider now the unavoidable first backup case (fourth step).
This case can be an improvement compared to the shared backup case
only if $s_2$ is disconnected from $t_2$ in $G_2$. We handle this
case by taking the best disjoint paths in $G_1$ without taking into
account which links belong (or not) to $b_1$. We get for the failure
probability of the second connection a value $q_1$ equal to the sum
of the failure probability $p_1$ and failure probabilities of bridge
links between $s_2$ and $t_2$ in $G_1$.

In the overlapped connection case, we can decouple the determination
of the two paths of the second connection. We recall that in this
case the only possible superpositions are between $p_1$ and $b_2$,
$b_1$ and $p_2$, and $p_2$ and $b_2$. The establishment of the
primary $p_2$ does not interfere with the choice of $b_2$, hence
their establishment can be done independently. Indeed, according to
the minimization expression (\ref{eq:C1}), in this case the failure
probability of the second connection can be expressed by $P_f (p_2)
+ P_f (p_1 \cap b_2)$. Thus, we can take for $p_2$ the shortest path
from $s_2$ to $t_2$ in $G_1$ (the weights are the failure
probabilities of the links). The choice of $b_2$ corresponds then to
a shortest path from $s_2$ to $t_2$ in a graph $G'_1$ where the
links of $b_1$ have been deleted from $G$ and the only positive
weight links are the links of $p_1$ with their real failure
probabilities (fifth step). Finally, we compare the solution between
the two last different cases and we return its corresponding primary
and backup paths (sixth step).

We proceed to prove the polynomial complexity of Algorithm SCA by
determining the complexity time of each successive step. The first
step can be computed in $O(n \cdot m)$ steps: if it exists, a bridge
link should be on the shortest path between $s_2$ and $t_2$ that can
be computed in $O(m+n \log n)$ time using Dijkstra's algorithm
\cite{dijkstra}. For each link of the shortest path (less than $n$),
we determine if it corresponds to a bridge link by removing it and
by observing if there is some connectivity in the resulting graph
between $s_2$ and $t_2$ ($O(m)$ steps). The second step can be done
in $O(m)$ steps: after the removal of all the bridge links found in
the first step, it checks only the connectivity of the different
components.

The third step can be done in $O(m \log_{1+\frac{m}{n}} n)$ steps:
we can improve the complexity time of \cite{tunable} by duplicating
all the bridge links found in the first step and computing in the
resulting graph the two link-disjoint paths according to
Suurballe-Tarjan's algorithm \cite{suurballe}, which runs in $O(m
\log_{1+\frac{m}{n}} n)$ number of steps. The fourth step has the
same complexity since the verification of a shared connected
component in $G_1$ between $s_2$ and $t_2$ can be done with
Dijkstra's algorithm, and the computation of the improved algorithm
inspired from \cite{tunable} has been established in the third step.
At last, only successive shortest path computations are done in the
fifth step, in $O(m n \log n)$ time.}
{ We describe here a sketch of the proof; the full details are in
\cite{tr}. We consider the three generic cases (shared backup,
unavoidable first backup, overlapped connections) that corresponds
to the optimization expression (\ref{eq:C1}), as described above,
and we analyze the optimality of the solution obtained by Algorithm
SCA under each case, as well as the corresponding computational
complexity.

Consider first the shared backup case: $p_1$ has no
superposition with other paths, while $b_1$ potentially has
some superposition with only $b_2$. The first step of SCA
indicates when no solution can be found in the shared backup
case, and thus the other cases should be analyzed. Otherwise,
by using the decomposition described in the second step of SCA,
we can prove that we can find two link-disjoint paths from
$s_2$ to $t_2$ that use a minimal number of bridge links (third
step of SCA). As indicated by expression (3), we can compute
the minimal failure probability of a single connection in $G$
as the sum of the bridge links between the source and the
destination.

In addition, this failure probability is minimal amongst the
different cases: in the unavoidable first backup case, the same
links have to be shared by the paths of the second connection
since they are also bridge links in $G_1$. In the overlapped
connection case, the failure probability of the second
connection includes the one of the primary path $p_2$
and, by construction, $p_2$ uses all the bridge links
between $s_2$ and $t_2$ in $G_1$. As a consequence, its failure
probability is superior to that obtained by the shared backup
case.

We consider now the unavoidable first backup case, if there is no
solution for the shared backup case. The optimal solution in this
case can be obtained by using the algorithm of \cite{tunable}
in $G_1$, as done in the fourth step. In the overlapped
connection case, the failure probability of the second connection
can be expressed by $P_f (p_2) + P_f (p_1 \cap b_2)$. Thus, we can
decouple the selection of the paths $p_2$ and $b_2$ (fifth step).
Finally, we compare between the solutions of the two last different
cases, and return the best primary and backup paths (sixth step).

We proceed to analyze the running time of Algorithm SCA. The first
step can be computed in $O(n \cdot m)$ steps: if it exists, a bridge
link should be on the shortest path between $s_2$ and $t_2$ that can
be computed in $O(m+n \log n)$ time using Dijkstra's algorithm
\cite{dijkstra}. For each link of the shortest path (less than $n$),
we determine if it corresponds to a bridge link by removing it and
by observing if there is a path in the resulting graph between $s_2$
and $t_2$. The second step can be done in $O(m)$ steps: it just
checks the connectivity of the different components. The third step
can be done in $O(m \log_{1+\frac{m}{n}} n)$ steps: we can improve
the complexity time of \cite{tunable} by duplicating all the bridge
links found in the first step and computing in the resulting graph
the two link-disjoint paths according to Suurballe-Tarjan's
algorithm \cite{suurballe}, which runs in $O(m \log_{1+\frac{m}{n}}
n)$ steps. The fourth step has the same complexity since it uses the
same algorithms. Finally, only two successive shortest path
computations are done in the fifth step, in $O(m+n \log n)$ time. }
\proofbox

In Problem 2CP-2, the primary path $p_1$ has been established
already, and a backup path $b_1$ and the two paths of the second
connection should be found such the failure probability of the
second connection is minimized while providing zero failure
probability to the first connection (i.e., obeying the constraint
$MCFP(c_1)=0$). As a consequence, the two paths of the first
connection must be link-disjoint. We proceed to prove the computational
intractability  of Problem 2CP-2.

\begin{mytheorem}
\label{thm: Th C} Problem 2CP-2 is NP-complete; moreover, it is
NP-hard to approximate its solution within a constant factor.
\ifthenelse{\boolean{TechReport}}
{\end{mytheorem}

\textbf{Proof:}

We shall employ a reduction to the shortest path with pairs of
forbidden nodes problem (termed {\em Problem SP-DPFN}). As shown in
\cite{faute}, the special case of Problem SP-DPFN where the degree
of the forbidden nodes is equal to two (termed {\em Problem
SP-DPFN2}) is Min-PB\cite{kann}; the latter is a class of
minimization problems whose objective functions are bounded by a
polynomial in the size of the input. Problems of this class  are
hard to approximate, which implies that Problem SP-DPFN2 is
NP-complete and it is NP-hard to approximate its solution within a
constant factor.

\begin{figure}
\centering
\includegraphics[width=\smallgraphsize  \textwidth]{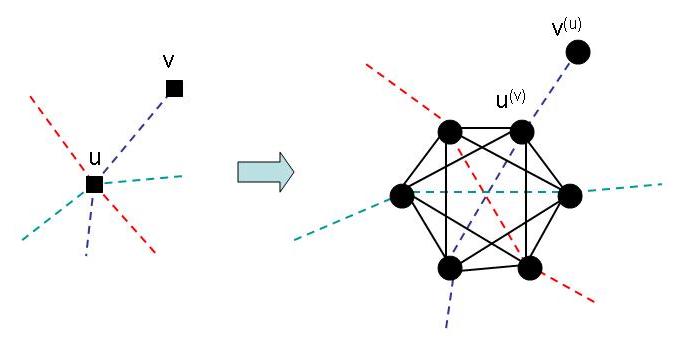}
\caption{The clique transformation.} \label{fig:clique}
\end{figure}

We consider an instance $(G,w,s,t,W)$ where $G=(V,E)$ is the graph
of  the instance, $w$ its weight function, $s$ and $t$ are,
respectively, the source and the destination, and $W$ is the
collection of the disjoint pairs of forbidden nodes. We apply the
following {\em clique transformation} to $G$. Each node with a
degree larger than two is transformed into a clique whose number of
nodes equals the degree of the original node and the internal links
have zero weight (see Figure~\ref{fig:clique}). We obtain a new
graph called $G_1=(V_1,E_1)$. For each clique in $G_1$representing a
node $x_i$ in $V$ and each link $(x_i,x_j)$ in $E$, we call
$x_i^{(x_j)}$ the node of this clique which is linked to a node of
the clique in $G_1$ representing $x_j$ ($x_j \in V$).
This transformation can be done within a polynomial number of steps.
Even in the worst case of a clique $K_n$ (with $n$ nodes and
$m=\frac{n(n-1)}{2}$ links) as the original graph, after the clique
transformation, the obtained graph will be composed of $n(n-1)$
nodes and $\frac{n(n-1)^2}{2}$ links, since each node has a degree
$n-1$.

The subproblem of SP-DPFN2 where the graph of the instance has
been transformed by a clique transformation remains
intractable. Indeed, an immediate correspondence can be found
between paths of both instances. In addition, even if we assume
that $G$ is Eulerian, this restriction does not modify the
tractability of the SP-DPFN2 problem for an instance with $G$
as the graph.

We consider an instance $(G,w,s,t,W)$ of the SP-DPFN2 problem, such
that the original undirected graph $G$ is Eulerian and has been
transformed to $G_1=(E_1,V_1)$ by the clique transformation. $s$ and
$t$ are, respectively, the source and the destination of the
shortest path. The collection of disjoint forbidden pairs of nodes
$W$ includes $\{(x_1,y_1),..,(x_l,y_l) \}$. We build from this
instance a new instance to Problem 2CP-2 as follows. We add two
nodes $s'$ and $t'$ to $V_1$ and we link them to the source $s$ and
to the destination $t$ of the shortest path (i.e. we add $(s',s)$
and $(t',t)$ to $E_1$). We transform each node of the different
pairs of forbidden nodes according to the following {\em ``bow tie''
transformation}.
\begin{figure}
\centering
\includegraphics[width=\smallgraphsize  \textwidth]{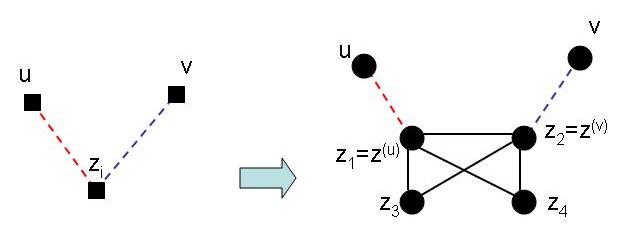}
\caption{The bow tie transformation.} \label{fig:bow}
\end{figure}
As described in Figure~\ref{fig:bow}, with this transformation a
node $z_i$ with a degree two is split into four interlinked nodes
$\{z_i^1,z_i^2,z_i^3,z_i^4 \}$, with additional links of zero weight
$\{(z_i^1,z_i^3),(z_i^2,z_i^4),(z_i^1,z_i^4),(z_i^2,z_i^3) \}$. In
addition, both $z_i^1=z_i^{(u)}$ and $z_i^2=z_i^{(v)}$ are endpoints
of the respective outgoing links $(z_i,u)$ and $(z_i,v)$. (We note
that $z_i^1$ and $z_i^2$ are linked by a positively weighted link.)
We will see that this link is necessarily used by the primary path
$p_1$. Beyond this transformation, we add a new node set
$V_2=\{v_1,..,v_{l+1}\}$ and the link set
$F_2=\{(s,v_1),(v_i,x_i^3),(v_i,y_i^3),(x_i^4,v_{i+1}),(y_i^4,v_{i+1}),(v_{l+1},t),
1\leq i \leq l\}$ such that each link has a zero weight. Finally, we
take $s_1=s_2=s$ and $t_1=t_2=t$.
\begin{figure}
\centering
\includegraphics[width=\smallgraphsize  \textwidth]{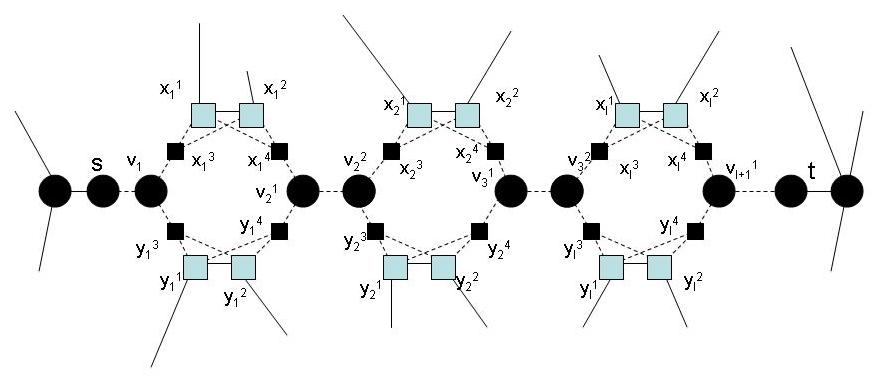}
\caption{The reduction for Problem 2CP-2.} \label{fig:2cp3}
\end{figure}
Figure~\ref{fig:2cp3} sums up this new instance composed of a graph
$G'=(V',E')$. For ease of drawing, the regular nodes of $G$ are not
included. We assume that $p_1$ has been chosen in order to go
through all the nodes in the original graph $G$ (which is possible
because $G$ is Eulerian). Thus, if $p_1=(r_1,..,r_i,..,r_j,..,r_n)$
such that $r_i$ is assumed to be a member of a pair of forbidden
nodes and $r_j$ a regular node, the primary path $p_1$ will be
composed of $r_i^{(r_{i-1})}-r_i^{(r_{i+1})}$ instead of $r_i$, and
of $r_j^{(r_{j-1})}$ and $r_j^{(r_{j+1})}$ instead of $r_j$.

Assume we have an algorithm that solves Problem 2CP-2.  In addition,
instead of a uniform weight distribution, assume that the weight of
links of type $(z_i^1,z_i^2)$ are identical and equal to $m$ times
the uniform weight of the other weighted links. Since $s$ and $t$
have a node degree of two, the shared backup case is not relevant
here. The optimal solution is thus due to the unavoidable first
backup case or to the overlapped connection case. In the first case,
the three paths $b_1, p_2$ and $b_2$ can only use the additional
zero weight links, since these paths should be link-disjoint with
$p_1$. Thus, by taking an identical path from $s$ to $t$ through
these zero weight links, we get that the failure probability of the
second connection is equal to $P_f (p_1)$. Considering the
overlapped connection case, for the same reasons $b_1$ and $p_2$ are
allowed to use only zero weight links. Since in this case the
failure probability of the second connection is $P_f(p_1 \cap b_2)$
(we recall that $P_f(p_2)=0$), the minimal failure probability of
the second connection reflects the number of links of kind
$(z_i^1,z_i^2)$ used because of their relatively large weight.
Moreover, the backup path $b_2$ only uses links present in the
original graph $G$ except maybe some links of the bow ties. The
backup path may go through $z_i^1$ and $z_i^2$ without using the
link $(z_i^1,z_i^2)$ only if both the backup path $b_1$ and the
primary path $p_2$ do not go neither through $z_i^1$ nor $z_i^2$.
Thus, by evaluating the minimal failure probability of the second
connection, we are able to deduce if a link of kind $(z_i^1,z_i^2)$
has been used. If no such link has been used, this backup path $b_2$
is the shortest path in $G$ from $s$ to $t$ that does not use both
nodes of any pair of forbidden nodes. Thus this path would solve the
SP-DPFN2 problem for this instance.

Conversely, from any solution to the SP-DPFN2 problem with the
original instance we can construct a solution to Problem 2CP-2, as
follows. We consider for the overlapped connection case the backup
path $b_2$ that corresponds to the respective path in $G'$, and for
$b_1=p_2$ a path that is composed of zero weight links and is link
disjoint with $b_1$ (such a path exists, since the initial shortest
path does not use both nodes of a pair of forbidden nodes). As we
have seen, this combination leads to the minimal failure probability
for the second connection in the overlapped case, and the solution
is obtained by comparing it to the failure probability of the
solution obtained in the unavoidable first backup case (i.e.
$P_f(p_1)$). \proofbox
}
{\proofbox
\end{mytheorem}}

We are ready to present our heuristic scheme for solving Problem
2CP-2, termed {\em Algorithm 2CP-2A}. The basic idea of this
algorithm is to return the best solutions in the shared backup case
(first step of Algorithm 2CP-2A) and the unavoidable first backup
case (second step of Algorithm 2CP-2A). We do not consider the
solutions that respect the constraints of the overlapped connection
case, because, as indicated in the proof of Theorem~\ref{thm: Th C}
(presented in \cite{tr}), they are the source of the intractability
of Problem 2CP-2. In addition, Algorithm 2CP-2A always returns a
(not necessarily optimal) solution if there exists at least one
solution $\{(b_1,p_2,b_2)\}$ that respects the constraints of the
overlapped connection case. For example, the solution with the same
paths except that the second backup path is chosen identical to
$p_2$, respects the constraints of the unavoidable first backup case
since both paths of the second connection are link-disjoint with the
first primary path.

{\bf\em Algorithm 2CP-2A:}

\textbf{First step.} In the graph $G_1$ representing the
original graph $G$ without the links of $p_1$, find two
link-disjoint paths, one from $s_1$ to $t_1$, the other from
$s_2$ to $t_2$, using the two link-disjoint path algorithm
of \cite{tholey}. If such paths are found, assign for
$b_1$ the path from $s_1$ to $t_1$. Then, given the two paths
of the first connection, assign to the second connection the
paths returned by Algorithm SCA.

\textbf{Second step.} If no such paths are found, assign to $b_1$
the shortest path from $s_1$ to $t_1$, and assign to the second
connection the paths obtained by \cite{tunable} in $G_1$.

In Section V we shall present the results of a simulation study
on the performance of Algorithm 2CP-2A.

We proceed to consider Problem 2CP-3. Here, four paths should be
found such that the first connection is fully reliable (namely,
$MCFP(C_1)=0$), and the paths of the second connection minimize its
failure probability. Similarly to Problem 2CP-2, we prove the
hardness of Problem 2CP-3 and, based on the insight gained through this
analysis, we propose
\ifthenelse{\boolean{TechReport}}
{heuristic solution schemes.}
{a heuristic solution scheme.
}

\begin{mytheorem}
\label{thm: Th D} Problem 2CP-3 is NP-complete; moreover, it is
NP-hard to approximate its solution within a constant factor.
\ifthenelse{\boolean{TechReport}}
{\end{mytheorem}

\textbf{Proof:}
We shall employ a reduction  from 2CP-3 to Problem
SP-DPFN2 (in its nodal version), similarly to the previous proof.
SP-DPFN2 remains Min-PB even in its nodal version\cite{faute}, which
implies that Problem SP-DPFN2 and thus Problem 2CP-3 is  NP-complete
and it is NP-hard to approximate it within a constant factor.

We consider the graph used in the proof of Theorem~\ref{thm: Th 1}.
In addition, we transform all the links in the original graph $G$
according to the following {\em duplicate link transformation}.
\begin{figure}
\centering
\includegraphics[width=\smallgraphsize  \textwidth]{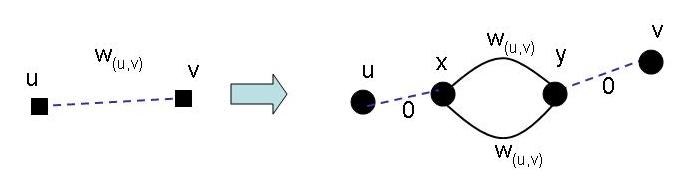}
\caption{The duplicate link transformation.} \label{fig:splitting}
\end{figure}
As depicted in Figure~\ref{fig:splitting}, with the duplicate link
transformation, we create, out of a link $(u,v)$, two additional
internal nodes $x$ and $y$, and four replacing links
$\{(u,x),(x,y),(x,y),(y,v) \}$ such that $w_{(u,x)}=w_{(y,v)}=0$ and
$w_{(x,y)}$ is equal to the previous $w_{(u,v)}$ for both links
$(x,y)$. Our goal is to determine the existence of a solution that
leads to a zero failure probability for the second connection. Since
the sources and the destinations are identical ($s_1=s_2=s$ and
$t_1=t_2=t$) and their degree is equal to two, there is no need to
consider the shared backup case similarly to the previous proof.
Indeed, finding paths that respect the constraints of the shared
backup case would imply that at least three link-disjoint paths
emanate from $s$ (namely, the two primary paths and the shared
backup path), which is impossible. In the unavoidable first backup
case, the total second connection failure probability includes the
path failure probability of the first primary path $p_1$ (since
$p_1$ and $b_1$ have no common links), which should thus equal zero.
Therefore, $p_1$ is allowed to use only the zero weighted links.
Thus, finding a zero failure probability for the second connection
is equivalent to finding a path in the original graph that includes
at most one node of each pair of forbidden nodes. In addition, by
minimizing the failure probability of the second connection, we aim
to find in the reduced problem the shortest path, which is a Min-PB
problem. In the overlapped connection case, the total failure
probability of the second connection includes the path failure
probability of the second primary path $p_2$. Similarly, this path
is allowed to use only zero weighted links, i.e. the links with zero
failure probability. Once again, if we take, for example, $b_1=p_2$
(it does not add any constraint), getting a zero failure probability
for the second connection means finding in the original graph, for
the first primary path $p_1$, a path including at most one node of
each pair of forbidden nodes.

Conversely, any solution to the SP-DPFN2 problem in its nodal
version leads to an optimal solution for Problem 2CP-3. Indeed, we
can take as $b_1, p_2$ and $b_2$ the same paths corresponding to the
solution of problem SP-DPFN2, such that the only unshared links
between $p_2$ and $b_2$ are the weighted links. In addition, we
select the first primary path from $s$ to $t$ such that only zero
weighted links are used. There exists at least one feasible path
because the solution of the SP-DPFN2 problem does not use both nodes
of each pair of forbidden nodes.\proofbox

We proceed to describe two heuristics for Problem 2CP-3, both of
which are evaluated by way of simulations, as described in Section
V. In both heuristics, we decouple Problem 2CP-3 into two problems.
For the first heuristic, termed {\em Algorithm 2CP-3A}, we first
assign the paths of the first connection, and, with these at hand,
we proceed to choose the paths of the second connection by employing
our (optimal) solution to Problem 2CP-1. In the second heuristic,
termed {\em Algorithm 2CP-3B}, we first choose the primary path of
the first connection, $p_1$, and, with this at hand, we choose the
other paths by employing our (heuristic) solution to Problem 2CP-2.
The first selection in both heuristics }
{\proofbox
\end{mytheorem}
We proceed to describe a heuristic scheme for Problem 2CP-3, termed
{\em Algorithm 2CP-3A}. In Section V we shall describe the results
of a simulation study of its performance. In this scheme, we first
select the paths of the first connection, and, with these at hand,
we proceed to choose the paths of the second connection by employing
our (optimal) solution to Problem 2CP-1. The first selection }
is done in a way that attempts to minimize interference with the
next connection; this is done by employing a heuristic rule similar
to that of  the MIRA scheme \cite{mira}, i.e., the link weights
attempt to reflect their ``criticality''. The choice of the weight
function is crucial for the success of such schemes, and several
functions are proposed in \cite{mira} that are correlated to the
link residual bandwidth. In order to avoid links that form part of
small link cuts (e.g. bridge links) in the choice of the paths, we
define a weight function as follows.
\begin{definition}
Given an undirected graph $G=(V,E)$ and a ``level'' $l$, the {\em
small-cut weight} of a link $e$ is:
\begin{equation}
w_e = \sum_{i=1}^l \frac{n_i}{i}
 \label{eq:sm}\
\end{equation}
where $n_i$ denotes the number of small cuts of size (exactly) $i$
including $e$, for $1\leq i\leq l$.
\end{definition}
In order to compute these small-cut weights, we use the polynomial
time algorithm of \cite{japan}, which computes, for an unweighted
graph, the list of the small cuts up to a specified level $l$.

{\bf\em Algorithm 2CP-3A:}

\textbf{First step.} Considering link weights according to the
small-cut weight function (Expression~\ref{eq:sm}), execute
Suurballe-Tarjan's algorithm\cite{suurballe}. Assign the two
link-disjoint paths that it outputs to the first connection.

\textbf{Second step.} Given the outcome of the first step, employ
Algorithm SCA in order to select the paths of the second connection.

\ifthenelse{\boolean{TechReport}}
{
{\em Algorithm 2CP-3B:}

\textbf{First step.} Select for $p_1$ the shortest path from $s_1$
to $t_1$ using any standard shortest-path algorithm (e.g. Dijkstra
\cite{dijkstra}), such that the link weight are obtained by the
small cut weight function.

\textbf{Second step.} Given the outcome of the first step, employ
Algorithm 2CP-2A in order to select the remaining three paths.

}
{}

\ifthenelse{\boolean{TechReport}}
{
\subsection{Unreliable First Connection}

\begin{mytheorem}
\label{thm: Th B} For a general value of $MCFP(c_1)$, Problem 2CP-1 is NP-hard.
\end{mytheorem}

\textbf{Proof:} The proof can be found in the Appendix
(Theorem~\ref{thm: Th B2}).\proofbox
}
{Before concluding our discussion on the establishment of two
connections, we note that in \cite{tr} we establish the
intractability of the following two extensions: the extension
of Problem 2CP-1 for a general (i.e., not
necessarily zero) value of $MCFP(c_1)$, and the variant of Problem 2CP-3 where both
connections need to be fully reliable (i.,e., $MCFP(c_2)$ is
also assumed to be zero). }

\ifthenelse{\boolean{TechReport}}
{\subsection{Reliable Second Connection in an Offline
Framework}

\begin{mytheorem}
\label{thm: Th B3} The problem of determining if the optimal
solution of Problem 2CP-3 results in two connections fully
reliable is computationally tractable.
\end{mytheorem}

\textbf{Proof:} The problem is equivalent to Problem 2-CESB, which
is proven in the Appendix to be computationally tractable
(Theorem~\ref{thm: Th 2}).\proofbox
}
{}

\section{Extension to $K$ Connections}

We proceed to consider Problem KCP, namely the problem of
establishing some $K$ connections, $K>2$. We have seen that, for
Problem 2CP, the computational tractability (or lack thereof)
of the problem variants depend on the level of reliability
demanded by the connections (namely,  whether they require full
reliability or not), as well as on whether an online or offline
framework is considered.

More specifically, we have seen that, in both online and
offline frameworks, Problem 2CP is intractable when not all the
connections are required to be fully reliable. One exception is
Problem 2CP-1, which considers an online framework in which the second
connection is not required to be fully reliable.
However, this
exception is quite artificial under an online framework when
the number of connections is unknown in advance: if
we tolerate the last connection to be unreliable, the
establishment of an (unknown) future request will become
intractable.

In view of the above findings, when moving to consider more than
two connections we restrict our attention  to fully reliable connections. We
begin with a brief overview of known results on the related $K$
disjoint paths problem. We then formulate an online variant of
Problem KCP, for which we establish a polynomial-time optimal
solution. We then discuss the tractability of several extensions of
the online problem. Finally, we show that the offline version is
also computationally tractable.

\subsection{The $K$ Disjoint Path Problem}
The $K$ disjoint path problem is, in fact, the problem of
establishing multiple connections, each consisting of only a single
path (i.e., no backup paths). Formally, the problem is as follows.
Given a collection of pairs of nodes $\{(s_1,t_1),..,(s_K,t_K) \}$
in a graph $G=(V,E)$, find $K$ mutually disjoint paths from $s_i$ to
$t_i$ for each $1\leq i\leq K$. For a general value of $K$, this
problem is known to be amongst the ``classic'' NP-Hard problems
\cite{karp}, both in its link-disjoint and node-disjoint versions.
On the other hand, if $K$ is fixed, i.e., $O(1)$, the problem is, in
principle, solvable in polynomial time, but the existing solutions
involve the computation of constant factors that are, in practice,
computationally prohibitive for  $K>2$ (yet they are practically
admissible for the case $K=2$)
\cite{tholey}. For  a fixed ($O(1)$) value of $K$, polynomial
solutions have been established, both for planar graphs
\cite{schrijver} and for general graphs \cite{seymour}.

\ifthenelse{\boolean{TechReport}}
{The tractability of the {\em weighted version} of the problem, i.e.
minimizing the total weight of all the $K$ disjoint paths (for a
fixed $K$ and given a weighted graph), is an open problem even for
$K=2$.}
{}

\subsection{The Online $K$ Connection Problem}
\subsubsection{Problem Formulation}
We focus on an online version of the $K$ connection problem,
for $K>2$. In this setting, at each step $K$, we check whether
we can establish a new connection between a source $s_K$ and a
destination $t_K$, given the paths allocated to the previous
connections at steps $1,2, \cdots , K-1$. At this stage, we can
distinguish between three disjoint sets of links, which
partition the original graph. The first set is the set of
\textit{free links} (that is, links not already used by any
paths), the set of \textit{backup links} (links only shared by
previous backup paths), and the set of \textit{the primary
links} (links only used by some primary path). For the new
connection, we allow its primary path to use only free links,
while its backup path is allowed to use free as well as backup
links.

\begin{definition}
For an instance $I=(G, E_1,E_2,E_3,s_k,t_k)$, where $G=(V,E)$ is an
unweighted undirected graph, $E_1, E_2$ and $E_3$ form a partition
of the link set $E$, corresponding to the link sets of free, backup
and primary links (respectively), and $(s_K,t_K)$ are two nodes of
$V$, \textit{the online $K$ connection problem} (in short, OKCP) is
the problem of finding two link-disjoint paths from $s_K$ to $t_K$
(``primary'' and ``backup'') such that none of the two paths uses
any primary links (i.e., links of $E_3$) and at most one of the
paths uses backup links (i.e., links of $E_2$).
\end{definition}

\subsubsection{Reduction for Problem OKCP}

\begin{lemma}
For each instance of Problem OKCP, there is a reduction R1 to an
instance where $s_K$ and $t_K$ are in the same two-link connected
component of the graph formed by the union of $E_1$ and $E_2$.
\end{lemma}

\textbf{Proof:} Given an instance $I_1=(G, E_1, E_2, E_3, s_K,
t_K)$, if $s_K$ and $t_K$ are not in the same two-link connected
component of the graph formed by the union of $E_1$ and $E_2$, by
Menger's Theorem\cite{menger} we are ensured that no two
link-disjoint paths from $s_K$ to $t_K$ can be found, even if we
allow both paths to use free and backup links.\proofbox

\begin{figure}
\centering
\includegraphics[width=\graphsize  \textwidth]{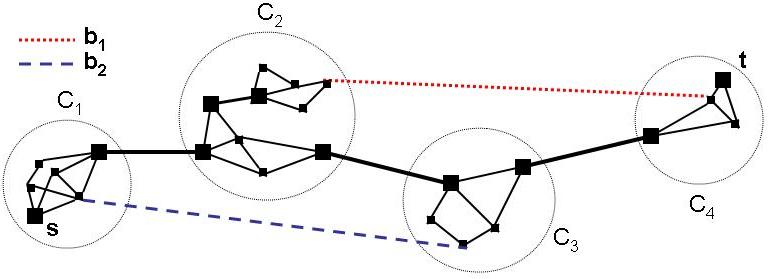}
\caption{Menger's Theorem is not enough.}
\label{fig:menger}
\end{figure}

The previous lemma gives a necessary condition for the existence of
a solution to Problem OKCP. Unfortunately, and in contrast to the
setting covered by Menger's Theorem, in our case the primary path
between $s_K$ and $t_K$ cannot use the backup links, hence the
condition is not sufficient, as demonstrated in
Figure~\ref{fig:menger}. This figure depicts the free links and the
two first backup paths $b_1$ and $b_2$ (the respective primary paths
are not drawn). Despite that $E_1 \cap E_2$ forms a two-link
connected graph, a new request between $s$ and $t$ must be declined.
We thus proceed with a more detailed study of the characterization
of the solution's feasibility.

\subsubsection{Leaps and Transitions} According to the previous
reductions, we can consider from now on that $E_1 \cup E_2$ forms a
two-link connected graph such that $E_1$ is one-link connected and
all its bridge links separate $s_K$ from $t_K$
\ifthenelse{\boolean{TechReport}}
{. The result is very similar to the one obtained after the
decomposition D1, except that here all the components are
necessarily two-link connected. In order to determine the existence
of a solution to Problem OKCP, we decompose the problem according to
the two-link connected components of $E_1$. }
{in different two-link connected components. }
For each component, we analyze which are the link-disjoint paths
that may be part of a solution (if such a solution exists), and
then we determine if the instance is solvable. We shall need the
following definitions.

\begin{definition}
Given a  pair of nodes $\{u,v\}$, a source $s$ and a destination
$t$, we say that a path $p$ from $s$ to $t$ has a \textit{leap}
$\{u,v\}$ if $p$ is composed of two link-disjoint paths
corresponding either to a path from $s$ to $u$ and another path from
$v$ to $t$, or to a path from $s$ to $v$ and another path from $u$
to $t$.
\end{definition}

\begin{definition}
Given a two-link connected component of free links, including four
nodes $\{s_1, t_1, s_2, t_2 \}$ and a set $B$ of nodes of this
component connected to a backup link, a leap set $\{
(u_1,v_1),..,(u_k,v_k) \}$ is called a \textit{feasible transition}
if all the nodes of the leap set are in $B$, and we have two link
disjoint paths, one from $s_1$ to $t_1$, and the other from $s_2$ to
$t_2$ with the leaps of the leap set.
\end{definition}

\ifthenelse{\boolean{TechReport}}
{Given a two-link connected component $C$ of free links with four
nodes $\{s_1, t_1, s_2, t_2 \}$, we specify in the Appendix the
conditions on the instance $(C,s_1,t_1, s_2, t_2)$ }
{\cite{thomassen} has specified the conditions on a graph}
for admitting a transition without any leap. In addition, we can use
a $K$ link-disjoint path algorithm\cite{seymour} to determine the
existence of a transition with $K-1$ leaps. Thus we can list all the
possible transitions for each component, for any number of leaps. We
proceed with a specification of the algorithm that solves Problem
OKCP.

\subsubsection{Algorithm RCA} We present here the {\em Reliable
Connection Addition} algorithm ({\em Algorithm RCA}), that returns a
solution to Problem OKCP if there is one, and indicates its absence
otherwise.

\textbf{First step.} Let $G_1$ be the graph of free links. Denote by
$C_1$ the connected component in $G_1$ to which $s_1$ belongs. If
$t_1$ is not located in this component, return no solution.
Otherwise, assign all the free links not in $C_1$ to the backup link
set $E_2$.

\textbf{Second step.} If $G_1$ is a two-link connected graph, we can
find two link-disjoint paths from $s_K$ to $t_K$ using only links of
$E_1$.
Else if $E_1 \cap E_2$ contains a bridge link that separates $s_K$
from $t_K$, return no solution (Reduction R1).

\textbf{Third step.} Consider all the possible combinations of
backup link sets with no redundancy by progressively incrementing
the maximal number $N_1$ of leaps by transition, as follows.

$N_1 \leftarrow 0$
\begin{algorithmic}
\WHILE{$N_1<K$}
\STATE
compute all the possible transitions with at most $N_1$ leaps;
observe if there is at least one combination that
goes from the first component to the last one, such that all
the transitions are feasible and in different components, and such that for
each leap $(u,v)$ there is a corresponding path in another
component from the backup link set of $u$ to the one of $v$.
If a feasible combination is found, go to the fourth step, else
increment $N_1 \leftarrow N_1+1$ \ENDWHILE
\end{algorithmic}
Stop when $N_1=K$ and return no solution.

\textbf{Fourth step.} Compute the respective transitions and return
the feasible primary and backup paths.

\begin{mytheorem}
\label{thm: Th E} Algorithm RCA solves Problem OKCP within
$O(max(n^4,n^{K-2}))$ steps.
\end{mytheorem}

\textbf{Proof:} The first step verifies that $s_K$ and $t_K$
are in the same connected component $C_1$ for an eventual path
$p_K$. Otherwise, no feasible solution can be found. The
assignment of the free links not in $C_1$ to $E_2$ does not
modify the number of solutions since no primary path $p_K$ will
be able to use these free links.
We use Lemma 1 to ensure that no
bridge links separate $s_K$ from $t_K$ in the graph formed by
$E_1 \cap E_2$. We recall that after this assignment, the
number of connected components of $E_2$ is inferior or equal to
the number of backup paths, i.e. $K-1$, because each backup set
has at least a backup link that a backup path uses with an
endpoint in $G$.

In the third step,
proceeding by induction on the maximal number of leaps by
transition, we consider all the possible combinations of backup link
sets that could lead to a solution. Indeed, if a solution exists,
its maximal number of leaps for a transition cannot be more than
$K-1$, since a backup link set cannot be used twice for a given
transition. Thus, by Algorithm RCA, a solution is found only if one
exists, and the solution returned is such that the maximal number of
leaps by transition is minimal. In the fourth step, the computation
of the primary and backup paths that use these transitions is done
in order to return paths that would form a feasible $K$th
connection.

The number of considered combinations remains polynomial since the
length of each cannot be greater than the bounded number of link
backup sets. Thus, the algorithm runs in polynomial time.
\ifthenelse{\boolean{TechReport}}
{We proceed to compute the running time of Algorithm RCA. The first
step needs $O(m)$ to verify that $s_1$ and $t_1$ are in the same
shared connected component. The second step runs in $O(n \cdot m)$
number of steps, since it looks for the bridge links between $s_K$
and $t_K$ both in graphs $G_1$ and $E_1 \cap E_2$. As seen in the
proof of Theorem~\ref{thm: Th A}, bridge links between separating
two nodes can be found in $O(n \cdot m)$ steps. In addition, if no
bridge edge is found between $s_K$ and $t_K$ in $G_1$, we use
Suurballe-Tarjan's algorithm that runs in $O(m \log_{1+\frac{m}{n}}
n)$ time when returning the solution.

In the third step, the number of combinations is constant regarding
to $n$ since it depends only on the number of backup link sets,
which is bounded by $K-1$. For each component $C_i$, we can compute
all its possible transitions by ignoring the other components and in
$O(n^4)$ steps: we remove all the edges (including the bridge edges
separating $s_K$ from $t_K$) in $G_1$ which are not part of the
component and all the backup link sets that have no common node with
$C_i$. We denote by $a_i$ and $b_i$ the two nodes of $C_i$ that
should be linked only by free links in order to be part of the
primary path. In addition, in $E_2$, we replace each set $S$ of the
remaining backup link sets by a single node $x_S$ linked to all the
common nodes of this set with $C_i$. At last, we transform all the
nodes endpoints of free links according to the clique transformation
(see the proof of Theorem~\ref{thm: Th C} for its definition). This
transformation can be done in $O(m)$ steps: we first compute the
degree of each node in order to create a corresponding number of
nodes in the transformed graph, and then we assign the linkage
according to each link in the original graph. Thanks to the clique
transformation, the computation of link-disjoint paths in the
original graph is equivalent to the computation of node-disjoint
paths in the transformed graph by an immediate correspondence.
Moreover, since we have not transformed the added nodes $\{x_S \}$,
the computation of node-disjoint paths in the resulting graph
prevents us from receiving a primary path using backup links.
Indeed, for a specified component and for a combination of backup
link sets $\{ (S_1,..,S_{2r}), r \leq N_1 \}$, we use the disjoint
path algorithm \cite{seymour} (which runs in $O(n^3)$ time) in its
nodal version to find the $r+1$ node-disjoint paths that would
compose the transition.
These paths are $\{(a_i,b_i), (x_{2l-1}l,x_{2l}) | 1 \leq l \leq r
\}$. Since we do the computation of all the transitions for $O(n)$
components, the resulting number steps is $O(n^4)$.

Once we have listed all the transitions with less than $N_1$ leaps
for each component, we observe if a given combination $\{
(S_1,..,S_{j})\}$ can result to feasible primary and backup paths.
We denote $c$ the number of components ($c \leq n$). This
verification is time-consuming and can be done in $O(n^{K-2})$
steps: we consider all the possibilities where $(S_i, S_{i+1}), 1
\leq i \leq j-1$ is in the $h^(i)$-th component, $1 \leq h^(i) \leq
c$. For this possibility, we verify if in each of the $j-1$ chosen
components there is a feasible transition composed exactly of these
backup link sets. Since we know that $S_1$ and $S_j$ should have a
common node with the first and last component respectively, the
possibility is characterized by the chosen components $\{h^(i), 1
\leq i \leq j-1 \}$ and the number of possibilities is $O(n^(K-2))$
because $j$ is inferior or equal to $K-1$ (the maximal number of
backup link sets).

When such a possibility is found, the fourth step computes the
primary and backup paths according to this result.}
{The complete analysis of the running time can be found in \cite{tr}.}
\proofbox

\ifthenelse{\boolean{TechReport}}
{\subsubsection{Heuristic Improvements}

Beyond Algorithm RCA, heuristic improvements can be done in order to
find faster a solution. These improvements can be a reduction of the
graph of free links, or an interference minimization consideration,
or a tradeoff between quality of the solution and computation time.

\begin{lemma}
For each instance of Problem OKCP, there is a reduction R2 to an
instance where $E_1$ is exactly a one-link connected subgraph
including $s_K$ and $t_K$ in separated two-link connected
components, such that all the bridge links in $E_1$ separate $s_K$
from $t_K$.
\end{lemma}

\textbf{Proof:} We consider an instance $I_1=(G, E_1, E_2, E_3, s_K,
t_K)$. $s_K$ and $t_K$ should be in a common connected component of
$E_1$ otherwise no feasible primary path could link $s_K$ to $t_K$
with only free links. Moreover, the primary path is not able to
access the other connected components of $E_1$, while the backup
path may be able to do so. Consequently, we can assign the links of
the other components of $E_1$ to the backup link set $E_2$. We can
then reduce $E_2$ to only its connected subgraphs having a node in
common with $E_1$ (the other backup links are not accessible to the
backup path). We observe that the number of connected subgraphs of
$E_2$ is bounded by the number of connections already established
(i.e. $K-1$): each backup path cannot be in more than a single
connected component, since it forms a connected path. In addition,
each connected subgraph of $E_2$ includes at least a backup path,
otherwise it would have no common node with $E_1$. Moreover, $s_K$
and $t_K$ are in separated two-link connected components of $E_1$:
otherwise, by Menger's Theorem\cite{menger}, we could (easily) find
two link-disjoint paths from $s_K$ to $t_K$ using only free links,
which would solve Problem OKCP for this instance.

We can reduce $E_1$ such that the only bridge links are the links
separating $s_K$ from $t_K$. Indeed we consider another bridge link
$e$ that separates $E_1$ into three edge sets $e\cup F\cup F'$, such
that $F$ and $F'$ are connected and $s_K$ and $t_K$ are in $F$.
Without modifying the feasibility of the solution, we can assign $e$
and the links in $F'$ either to the backup link set if $F'$ and
$E_2$ have at least a common endpoint, or to $E_3$ otherwise. Indeed
no primary path can use these links without using $e$ twice, and
this is also the case for the backup path if $F'$ and $E_2$ have no
common endpoint. In addition, since we have not created new
connected components, the number of backup link connected components
is still bounded by $K-1$.\proofbox

As in the case of two
connections, we can heuristically improve the success probability of
future connections by minimizing the ``interference'' with the
(unknown) future connections. Thus, we should assign weights to the
links such that they reflect their ``criticality''. Accordingly, we
modify the last step of the algorithm as follows: we compute the
primary path as before, and, given the primary path, we minimize the
weight of the backup path from $s_K$ to $t_K$ in the graph formed by
the union of $E_1$ with $E_2$. The weight of a link in $E_1$
corresponds to its criticality as defined by the small cut weight
function in the graph of free links. The weights of links in $E_2$
is zero since they do not interfere with future primary paths.

In addition, there is an inherent tradeoff between quality of the
solution and computation time. Indeed, the computation time of this
algorithm is not polynomial in the bounded number of backup link
sets. Moreover, we cannot hope to find such an algorithm: the
respective Problem OKCP for a general $K$ can be easily reduced to
Problem 2DPFL, which is proven in the Appendix to be NP-hard. Thus,
it may be interesting to keep a rather small number of backup link
sets for each new connection establishment. For future connections,
the number of backup link sets does not increase only if the actual
backup path goes through at least one endpoint of previous backup
links. Consequently, if in the interference minimization of the
backup path, the chosen backup path increases the number of backup
link sets, we should compare its total weight to the minimal weight
of a backup path going through at least an endpoint of a backup link
(this comparison can be done in polynomial time). }
{}

\ifthenelse{\boolean{TechReport}}
{

\subsection{Intractable Extensions}

We proceed to indicate  the intractability of several extensions of
Problem KCP.

\subsubsection{Directed Networks}

For directed networks, multi-connection establishment problems are
generally NP-Hard. Indeed, even the establishment of two connections
without backup paths is already intractable: the two disjoint path
problem was shown to be intractable, both in its nodal as well as
its link versions\cite{fortune}.

\subsubsection{Unbounded $K$}

\begin{mytheorem}
\label{thm: Th F} For a general  (not $O(1)$) $K$, Problem KCP is
NP-complete. Moreover, it is NP-hard  to approximate it within a
constant factor.\end{mytheorem}

\textbf{Proof:} The proof can be found in the Appendix
(Theorem~\ref{thm: Th F2}). \proofbox
}
{

\subsection{Intractable Extensions}

We note that extensions of Problem KCP when either the
network is directed or when the number of connections is general
(i.e., $K$ is not fixed) are shown in \cite{tr} to be intractable.}

\subsection{The Offline $K$ Connection Problem}
\ifthenelse{\boolean{TechReport}}
{We proceed to consider the {\em offline} version of the $K$
connection problem. Here, the set of connections is given and the
objective is to establish all of them at once. We proceed with a
formal definition of the problem.

}
{}
\begin{definition}
For an instance $I=(G,s_1,t_1,..,s_k,t_k)$, with $G=(V,E)$ an
unweighted undirected graph,  and $\{(s_i,t_i),1 \leq i \leq K \}$
source-destination pairs of $K$ connections, \textit{the offline $K$
connection problem} (in short, OffKCP) is the problem of finding
$2K$ (``primary'' and ``backup'') paths from $s_i$ to $t_i$, for
$1\leq i \leq K$, such that no link can be shared by a primary path
and another (primary or backup) path.
\end{definition}

\begin{mytheorem}
\label{thm: Th G} For a fixed $K$, there is an optimal solution to
Problem OffKCP of polynomial time complexity.
\end{mytheorem}
\ifthenelse{\boolean{TechReport}}
{
\textbf{Proof:} The proof can be found in the Appendix
(Theorem~\ref{thm: Th G2}).\proofbox
}
{
\textbf{Proof:} See \cite{tr}. \proofbox
}

\section{Simulation Study}

We performed simulations in order to evaluate the efficiency of our
proposed heuristics, namely Algorithm 2CP-2A for Problem 2CP-2 and
Algorithms 2CP-3A
\ifthenelse{\boolean{TechReport}}
{and 2CP-3B }
{}
for Problem 2CP-3. These algorithms are compared with the optimal
values (obtained through a ``brute force'' solution with an
inherently prohibitive computational complexity), as well as with
the outcome of more straightforward heuristic alternatives.

First, we describe the various algorithms that have been compared
for each of the problems, namely Problems 2CP-2 and Problem 2CP-3.
We then detail the
network topology
and the
parameters chosen for the simulations.
Through these simulations, we show the efficiency of our heuristics
and discuss its causes.

\subsection{The Tested Algorithms}

\subsubsection{Problem 2CP-2}
We recall that, in this problem, the primary path of the first
connection, $p_1$, is given. In our experiments, it was chosen to be
the shortest path between $s_1$ and $t_1$.
We then compared three algorithms for selecting the first backup
path $b_1$ and the two paths, $p_2$ and $b_2$, of the second
connection. We recall that these paths need to be selected so as to
minimize the failure probability of the second connection. The three
algorithms were the following: a simple heuristic (termed {\em
Algorithm 2CP-2N}), our heuristic (Algorithm 2CP-2A), and an optimal
(brute force) algorithm ({\em 2CP-2BF}).

Algorithm 2CP-2A has been described in Section III. We detail here
the other two algorithms. In the simple heuristic 2CP-2N, the links
of $p_1$ are removed and we choose for $p_2$ the shortest path
between $s_2$ and $t_2$ , where the link weights are their failure
probabilities. If, after the removal of $p_2$, a path between $s_1$
and $t_1$ can be found, we choose it for the backup path $b_1$ and
we then select for the backup path $b_2$ a path that minimizes the
failure probability of the second connection given $p_1$, $p_2$ and
$b_1$; otherwise, we do not remove the links of $p_2$ and we select
for the backup path $b_1$ the shortest path from $s_1$ to $t_1$. We
then select $b_2$ such that the failure probability of the second
connection is minimized.

The optimal, brute-force algorithm 2CP-2BF computes all the possible
backup paths $b_1$ from $s_1$ to $t_1$ link-disjoint from $p_1$, and
it then applies Algorithm SCA for each pair of paths. This way,
Algorithm 2CP-2BF always returns paths that minimize the failure
probability of the second connection.

\subsubsection{Problem 2CP-3}
In Problem 2CP-3, no path is already established but $p_1$ is
constrained to be link-disjoint at least with $b_1$ and $p_2$. Our
goal is to minimize the failure probability of the second
connection. We compared
\ifthenelse{\boolean{TechReport}}
{five algorithms.

Similarly to Algorithm 2CP-3A, the first algorithm (termed {\em
2CP-3N}) select at first the two paths of the first connection with
Suurballe-Tarjan's algorithm\cite{suurballe}, and then the second
connection with Algorithm SCA. The only difference between  2CP-3N
and 2CP-3A is in terms of the weight function, which, for 2CP-3N, is
the link failure probability function and not the (more
sophisticated) small-cut weight function of 2CP-3A. Similarly to
2CP-3B, the second algorithm ({\em 2CP-3N2}) selects first $p_1$ as
the shortest path between $s_1$ and $t_1$ and computes the remaining
paths as in Algorithm 2CP-2A. Here too, the only difference between
2CP-3B and 2CP-3N2 is the weight function, which, for 2CP-3N2, is
the link failure probability (rather than the small-cut weight
function).

The difference of weight functions corresponds to a lack of
interference minimization for the two first algorithms. Indeed,
for each of the two first algorithms, the first path
establishment is done only according to the link failure
probability weights, and not according to the topological
location of the links in the network. Thus, a critical link for
the second connection (e.g., a link in a two-link cut that
separates $s_2$ from $t_2$) may be used even if this link is
essential to the second connection.

}
{three algorithms.

Similarly to Algorithm 2CP-3A, the first algorithm (termed {\em
2CP-3N}) selects at first the two paths of the first connection
with Suurballe-Tarjan's algorithm\cite{suurballe}, and then the
second connection with Algorithm SCA. The only difference
between  2CP-3N and 2CP-3A is in terms of the weight function,
which, for 2CP-3N, is the link failure probability function and
not the (more sophisticated) small-cut weight function of
2CP-3A.

The difference of weight functions corresponds to a lack of
interference minimization for Algorithm 2CP-3A. Indeed, the first
connection establishment is done only according to the link failure
probability weights, and not according to the topological location
of the links in the network. Thus, a critical link for the second
connection (e.g., a link in a two-link cut that separates $s_2$ from
$t_2$) may be used even if this link is essential to the second
connection.

}
In the simulations, only a fixed $l$ is considered for the
small-cut weight function. Otherwise, since there are $2^n$ cut
sets in a graph, we could get an excessive (exponential) number
of cut sets. A tradeoff between the computation time and the
precision of the small-cut weights leads us to choose $l=4$.

\subsection{Implementation and Results}
We randomly generate undirected networks according to the
power-law rule \cite{powerlaw}
(henceforth: {\em power-law networks}).
The link weights follow an exponential distribution (with a
parameter $\lambda=5$) and are then normalized such that their
sum equals one. In addition, we assume that a few links may not
have the necessary capacity requirement for a connection. Thus,
we choose for each link a probability $p=0.15$ not to have
enough capacity. The node degree follows a power-law
distribution $\beta x^{-\alpha}$, where $\alpha=2.1$ and
$\beta$ differs according to the number $n$ of nodes in the
network ($\beta=100$ for $n=12$ and $\beta=500$ for $n=100$).
These parameters (and in particular $\alpha$) have been chosen
following \cite{powerpara}.
In addition, the four nodes of the two source-destination pairs are
randomly (uniformly) chosen.



\begin{table}[t]
\begin{center}
\begin{tabular}{|c||c|c|c|}
\hline
Network & 2CP-2BF & 2CP-2A  & 2CP-2N\\
\hline
PL12 & 100 \% &  94.30 \% & 74.0 \% \\
PL100 & 100 \% &  95.21 \% & 72.9 \% \\
\hline
\end{tabular}
\newline\caption{Comparison of the efficiency of the algorithms for Problem 2CP-2.}
\label{tab:2CP3comp}
\end{center}
\end{table}
We denote by {\em PL12}
and {\em PL100} the simulations produced by the generation of
power-law networks with, respectively, $n=12$ nodes and 500.000
random networks, or $n=100$ nodes and 50.000 random networks.
Table~\ref{tab:2CP3comp} indicates the fraction of times each
algorithm finds a solution with a minimal value for the failure
probability of the second connection in Problem 2CP-2. As expected,
we verify that Algorithm 2CP-2BF is always optimal.

\ifthenelse{\boolean{TechReport}}
{\begin{figure} \centering
\includegraphics[width=\biggergraphsize  \textwidth]{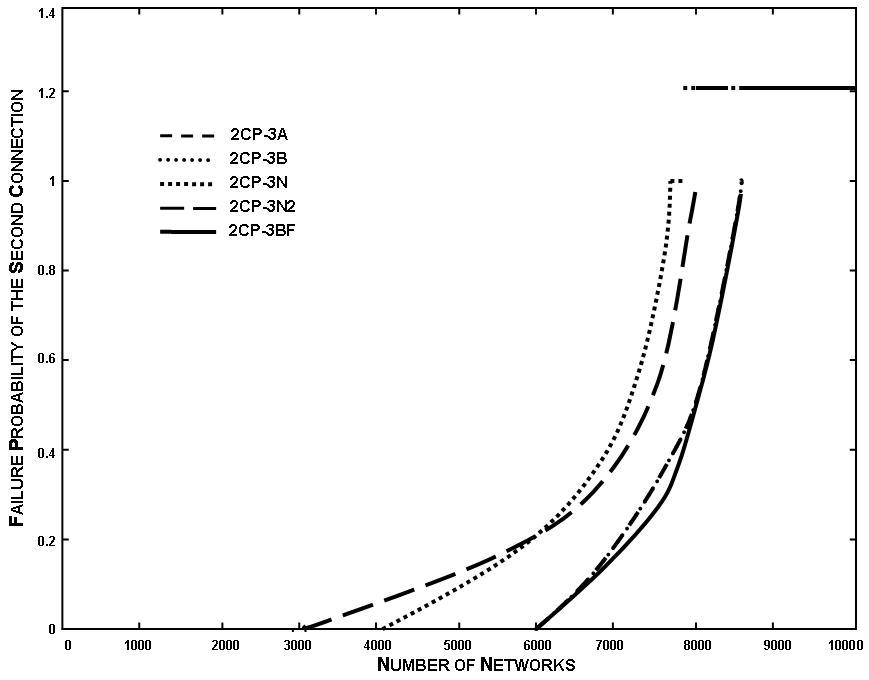}
}
{\begin{figure} \centering
\includegraphics[width=\biggergraphsize  \textwidth]{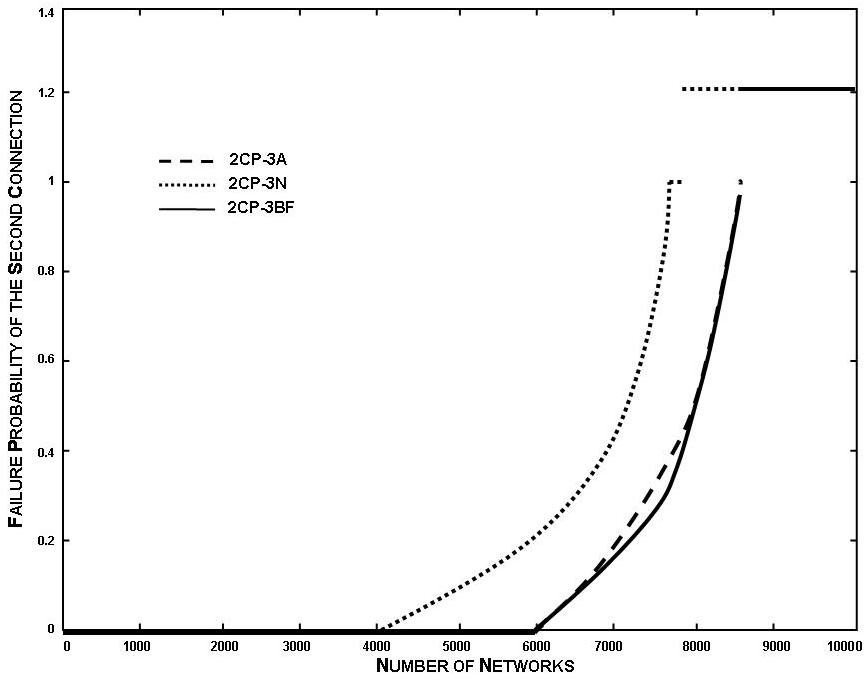}
}
\caption{Comparison of algorithms for Problem 2CP-3.}
\label{fig:simul}
\end{figure}

In Figure~\ref{fig:simul}, we consider Problem 2CP-3 and indicate
the distribution of the failure probabilities of the second
connection obtained after the simulations PL12 for the different
algorithms exposed above. Only 10.000 random power-law networks have
been generated for the network topology PL12 and none for PL100
because of the excessive computation time due to Algorithm 2CP-3BF.
When an algorithm finds a feasible solution, the failure probability
of the second connection is, obviously, under $1.0$. If no feasible
solution are found, we arbitrarily chose the value of this
probability to be equal to $1.2$ to be able to draw it on
Figure~\ref{fig:simul}.
Since the four
source-destination nodes are chosen randomly, in many occasions the
two requests do not interfere. This lack of interference is expected
to quickly diminish if the number of connections raises.
Table~\ref{tab:2CP4comp} shows the performance (i.e. the proportion
of times a minimal value is returned) for Algorithms 2CP-3BF, 2CP-3A
\ifthenelse{\boolean{TechReport}}
{2CP-3B, 2CP-3N and 2CP-3N2 for the network topology PL12.
\begin{table}[t]
\begin{center}
\begin{tabular}{|c||c|c|c|c|c|}
\hline
Network & 2CP-3BF & 2CP-3A  & 2CP-3B  & 2CP-3N  & 2CP-3N2\\
\hline
PL12 & 100 \% &  96.9 \% & 96.8 \% & 66.7 \% & 74.5 \% \\
\hline
\end{tabular}
\newline\caption{A second comparison of the algorithms for Problem 2CP-3.}
\label{tab:2CP4comp}
\end{center}
\end{table}
}
{and 2CP-3N for the network topology PL12.
\begin{table}[t]
\begin{center}
\begin{tabular}{|c||c|c|c|c|c|}
\hline
Network & 2CP-3BF & 2CP-3A  & 2CP-3N \\
\hline
PL12 & 100 \% &  96.9 \% & 66.7 \% \\
\hline
\end{tabular}
\newline\caption{A second comparison of the algorithms for Problem 2CP-3.}
\label{tab:2CP4comp}
\end{center}
\end{table}
}

\subsection{Discussion}

Consider first the simulations related to Problem 2CP-2. We observe
that the difference in efficiency between Algorithms 2CP-2A and
2CP-2N does not decrease with the increase of the network size.
Indeed, despite its increase, the network remains sufficiently
sparse to create interference between both connections. Furthermore,
we observe that Algorithm 2CP-2A is particulary efficient since in
most cases it returns paths with the same (optimal) quality obtained
by Algorithm 2CP-2BF.

Similar observations can be derived for Problem 2CP-3.
\ifthenelse{\boolean{TechReport}}
{Indeed, all
the algorithms are relatively efficient, finding in most cases the
optimal value returned by Algorithm 2CP-3BF.
In addition, we note that there is only a small difference
between the efficiency of 2CP-3A and 2CP-3B, as well as between
2CP-3N and 2CP-3N2. It means that in the fist step of the
algorithm, establishing only the primary path $p_1$ or also the
backup path $b_1$ in the same step does not substantially
altere the performance of the algorithms. Conversely, the
efficiency of the interference minimization and also of our
choice of the weight function is visible. Indeed, Algorithms
2CP-3A and 2CP-3B outperform Algorithms 2CP-3N and 2CP-3N2, and
almost always returns optimal solutions (i.e., with a minimal
failure probability for the second connection).

}
{Indeed, the efficiency of the interference minimization and also of our
choice of the weight function is visible: Algorithm 2CP-3A
outperforms Algorithm 2CP-3N, and almost always returns an optimal
solution (i.e., with a minimal failure probability for the second
connection).

}

\section{Concluding Remarks}

This study analyzed the computational complexity of
establishing multiple connections in survivable networks. For
several problem variants, optimal solutions of acceptable
computational complexity have been established. Other variants
have been shown to be computationally intractable, and
heuristic schemes, based on the analysis of the respective
problems, have been proposed.

This study has been motivated by the fact that, in previous
work, only single-connection problems have led to optimal (and
computationally efficient) solutions. Our study started with the
basic, but yet already complex case where two connections
should be established. A differentiation was done between online
and offline frameworks, and between fully reliable and
unreliable connections.

We then extended our analysis for a fixed $K$ connection problem,
where each connection is $100 \%$ reliable. For online and offline
variants, optimal solutions of polynomial complexity have been
established. The intractability of some extensions of the problem
was shown.

While this study covered only several out of the many variants
of the problem of establishing multiple survivable connections,
we believe that its methodical analysis provides a starting
point for the rigorous exploration of this important class of
problems.


\ifthenelse{\boolean{TechReport}}
{
\section*{Acknowledgment}

The authors would like to thank...

}
{}



%

\ifthenelse{\boolean{TechReport}}
{\section{Appendix}

In the Appendix, we will first prove the intractability of both a
variant problem of 2CP when the first connection is already
established and is not required to be fully reliable, and of the two
disjoint path with forbidden link problem as well. We then prove the
tractability of the $K$ connection problem with shared backups,
focusing on the restricted case $K=2$ and then enlarging the result.
In addition, we prove the intractability of Problem 2CP when an
additional bandwidth constraint is added. Then we study the
potential restrictions for Problem 2CP. At last, we proceed to
characterize the transitions with zero and one leaps.

\subsection{Problem 2CP for an Unreliable First Connection}

\begin{definition}
Given a primary path $p_1$ and a backup path $b_1$ for the first
connection $c_1$ with possibly some superposition between the two
paths, {\em Problem 2CP-u} is the problem of finding paths $p_2$ and
$b_2$ for the second connection that minimize their failure
probability.
\end{definition}

The first connection is not required in Problem 2CP-u to be fully
reliable, hence $MCFP(c_1)$ may be larger than zero and the paths of
$c_1$ are not necessarily link-disjoint. As we shall show, this
problem is NP complete, and, moreover, it is NP-hard even to
approximate the solution within a constant factor. The minimization
of the failure probability of the second connection is:
\begin{eqnarray}
\min_{p_2,b_2} && P_f (p_2\cap b_2)+1_{b_1 \cap p_2 \neq \oslash,b_1 \cap b_2 \neq \oslash} \cdot P_f (p_1-b_1) \nonumber \\
&&+1_{b_1 \cap b_2 = \oslash, b_1 \cap p_2 \neq \oslash, p_1 \cap b_2 \neq \oslash} \cdot P_f (p_1 \cap b_2) \nonumber \\
s.t. &&p_1 \cap p_2 = \oslash \label{eq:C2}\
\end{eqnarray}

\begin{mytheorem}
\label{thm: Th B2} Problem 2CP-u is NP-hard.
\end{mytheorem}

\textbf{Proof:} We prove that, given the paths of the first
connection, determining if there is a feasible second connection
that admits a connection failure probability equal to zero is
already NP-hard. We reduce Problem 2CP-u to the two link-disjoint
path with forbidden link problem (Problem 2DPFL), which is proven in
the next section to be NP-hard. Given a graph $G=(V,E)$ with two
nodes $s$ and $t$ in $V,$ and a set $F=\{(u_1,v_1),..,(u_k,v_k) \}$
of forbidden links with $F\subseteq E$, Problem 2DPFL requires to
determine if there exist two link-disjoint paths from $s$ to $t$
such that at the most one of them is composed of links of $F$.

We consider an instance $(G,s,t,F)$ (same notations as above) of
Problem 2DPFL. From this instance we construct an instance
$(G',s_1,t_1,s_2,t_2,p_1,b_1)$ of Problem 2CP-u such that
$G'=(V,',E')$ with $V'=V$ and $E'=E\cup
\{(s,u_1),(u_l,v_l),(v_k,t),1\leq l\leq k \}\cup
\{(v_k,u_{k+1}),1\leq l\leq k-1 \}$ (multiple links are accepted in
our model) and a uniform weight distribution ($\forall e \in E, P_f
(e)=\frac{1}{m}$). We also fix $s_1=s_2=s$ and $t_1=t_2=t$. The path
$s - u_1 - v_1 .. u_k - v_k - t$ is chosen as primary path $p_1$,
and the path $s - u_1 - v_1 .. u_k - v_k - t$ corresponds to the
backup path $b_1$. However, the links $(u_l,v_l)$ present in both
paths are chosen different. In order to get a zero failure
probability for the second connection, only the shared backup case
should be considered. Indeed, the other two cases have necessarily a
positive failure probability for the second connection, including
either the failure probability of $p_1$ (the unavoidable first
backup case) or the failure probability of $p_2$ (the overlapped
connection case). Thus, in the shared backup case, the second
connection should be composed of link-disjoint paths such that only
one path (i.e. $b_2$ but not $p_2$) may have common links with $b_1$
and none of them with $p_1$. By construction, the problem precisely
corresponds to finding in $G$ two link-disjoint paths such that only
one of them may use links in $F$ (composed of links in $b_1$ not
used by $p_1$) and with $s=s_2$ and $t=t_2$.

Conversely, any solution to the 2DPFL problem where the links of
$p_1$ have been removed and the remaining links of $b_1$ correspond
to the forbidden links, is an optimal solution to Problem 2CP-u,
since in such a solution the failure probability of the second
connection is equal to zero. \proofbox

\subsection{The Two Disjoint Path with Forbidden Link Problem }
\begin{definition}
Given an undirected graph $G=(V,E)$, two nodes $s$ and $t$ in $V$
and a link subset $F \subseteq E$ (called the forbidden link set),
{\em the two link-disjoint path with forbidden link problem }
(termed {\em 2DPFL}) corresponds to finding two link-disjoint paths
from $s$ to $t$ such that at least one path includes no links in
$F$. In its weighted version, to each link $e\in E$ is assigned a
length $l_e$, and {\em the weighted two link-disjoint path with
forbidden link problem} (termed {\em weighted-2DPFL}) corresponds to
finding two link-disjoint paths from $s$ to $t$ such that at least
one path includes no links in $F$ and such that the total length
(the sum of the two path lengths) is minimized.

\end{definition}
The same problems accept a nodal version, where the disjoint paths
should be node disjoint, and the forbidden set is a node set in $V$.
The respective problems are called {\em the two node disjoint path
with forbidden node problem } (termed {\em 2DPFN}) and {\em the
weighted two node disjoint path with forbidden node problem }
(termed {\em weighted-2DPFN}).

\begin{mytheorem}
\label{thm: Th 1} Problem 2DPFN is NP-complete.
\end{mytheorem}

\textbf{Proof:} We show that there is a reduction to Problem 2DPFN
from Problem SP-DPFN2 within a polynomial number of steps. We remind
to the reader that in Problem SP-DPFN2, given an undirected graph
$G=(V,E)$, two nodes $s$ and $t$, and a collection of disjoint
forbidden pairs of nodes $W=\{(x_1,y_1),..,(x_l,y_l) \}$ such that
each node has a degree equal to two, Problem SP-DPFN2 consists in
finding a path from $s$ to $t$ that contains at the most one node of
each pair. This problem has been proved to be NP-Hard and even
Min-PB (see \cite{faute}).

We proceed to the reduction in the following way. We take an
instance $(G,s,t,W)$ of Problem SP-DPFN2, and we consider all the
$r$ links outgoing from $W$ ($r \leq 2l$). We replace the links
$W_e=\{(w_1^i,w_2^i), 1\leq i\leq r \}$ by adding a new intermediary
node $u^i$ in each link, getting the link set
$U_e=\{(w_1^i,u^i)\cup(u^i,w_2^i), 1\leq i\leq r \}$. We define $U$
by $U=\{u^1,..,u^r\}$. Then we add a new node set
$V_1=\{v_1,..,v_{l+1}\}$ and the link set
$F_1=\{(s,v_1),(v_i,x_i),(v_i,y_i),(x_i,v_{i+1}),(y_i,v_{i+1}),(v_{l+1},t),
1\leq i \leq l\}$.
\begin{figure}
\centering
\includegraphics[width=\smallgraphsize  \textwidth]{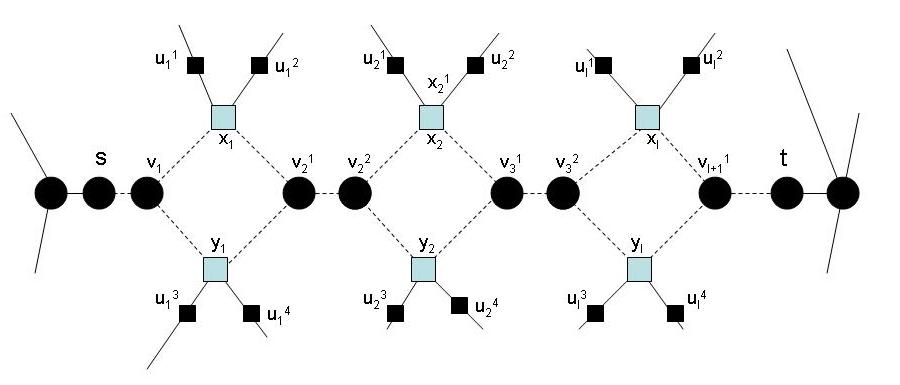}
\caption{The reduction from SP-DPFN2 to 2DPFN.} \label{fig:2dpfn}
\end{figure}
Figure~\ref{fig:2dpfn} depicts the obtained graph (for a better
clarity, the links of $G$ have not been drawn except the outgoing
links of $W$). Thus we have created a new instance to Problem 2DPFN
within a polynomial number of steps: the instance $(G',s',t',F')$ is
composed of a graph $G'=(V',E')$, such that $V'=V\cup U\cup V_1$ and
$E'=(E-W_e) \cup U_e \cup F_1$, and a forbidden node set
$F'=(V-W)\cup U$. The source and the destination are the same (i.e.
$s'=s$ and $t'=t$).

From any solution to Problem 2DPFN with the previous instance we can
construct a solution to Problem SP-DPFN2. Indeed, we can (easily)
verify that the feasible paths from $s$ to $t$ which have no nodes
in $F'$ are only those of the form
$(s,v_1,z_1,v_2,...,v_l,z_l,v_{l+1},t)$ where $z_i\in \{x_i,y_i\},
1\leq i\leq l$. Thus the other path is in fact a path from $s$ to
$t$, without any node of $V_1$ and any link of $F_1$, and with at
the most one node from each disjoint forbidden pair of nodes of $W$.
We can find an immediate correspondence between this path and a
solution in $G=(V,E)$ to Problem SP-DPFN2.

Conversely, if we have a solution to Problem SP-DPFN2 in $G=(V,E)$,
we can construct a path from $s$ to $t$ in $G'$ such that none of
its nodes is in $F'$. Then we can use for the other path the path
$(s,v_1,z_1,v_2,...,v_l,z_l,v_{l+1},t)$ such that $z_i\in
\{x_i,y_i\}, 1\leq i\leq l$. For each forbidden pair of nodes
$(x_i,y_i)$, $z_i$ is either the node not included in the first path
(there is always at least one since the first path is built from a
solution of SP-DPFN2), or $x_i$ if both are not included. These two
node disjoint paths represent a feasible solution to Problem 2DPFN.
\proofbox

\begin{corollary}
\label{cor: Co 1} The 2DPFL problem is NP-complete.
\end{corollary}

\textbf{Proof:} The outline of the proof for the link version of the
problem follows the same lines. We consider an instance of the
Problem SP-DPFN2 in order to find a linear reduction from it to
Problem 2DPFL. We split each node (of degree two) $z \in W$ into two
nodes, $z^1$ and $z^2$, such that each one is only an endpoint of
the two previous outgoing links from $z$ (if before the
transformation, we had two links $(z,u)$ and $(z,v)$, we have now
two links  $(z_1,u)$ and  $(z_2,v)$). In addition, we add the link set
$Z_e=\{(x_i^1,x_i^2),(y_i^1,y_i^2),1\leq i\leq l \}$, the node set
$V_1$ and the link set $F_1$ as defined in the previous proof. We
let the reader verify that in this way a valid reduction has been
built, and that this reduction can be done in polynomial time.
\proofbox

\begin{corollary}
Problem weighted-2DPFL is NP-complete and cannot be approximated
within a constant factor.
\end{corollary}

\textbf{Proof:} As proved in \cite{faute}, Problem SP-DPFN2 is Min
PB complete, which implies that it admits no approximation algorithm
within a constant factor. Thus, with the same previous reduction
where the link weights of the added links are equal to zero
(otherwise they are uniformly equal), we can deduce that the
weighted-2DPFL problem is also Min-PB complete, hence does not
accept an approximation algorithm within a constant factor.
\proofbox

\subsection{The $K$ Connection Problem with Shared Backups}
\begin{definition}
Given an undirected graph $G=(V,E)$ and $K$ pairs of nodes
($s_1$,$t_1$),..,($s_K$,$t_K$) in $V$, {\em The $K$ connection
establishment with shared backup path problem} (termed {\em K-CESB})
consists in finding $2K$ paths $\{ p_1,b_1,..,p_K,b_K \}$ connecting
their corresponding source-destination pair such that no link can be
common to different paths except if they are all backup paths.
\end{definition}

For example, for $K=2$, Problem {\em 2-CESB} consists in finding
four paths $p_1,b_1,p_2$ and $b_2$ connecting their corresponding
source-destination pair such that all pair of paths are link
disjoint except the two backup paths that may have common links.

\begin{mytheorem}
\label{thm: Th 2} Problem 2-CESB can be resolved in a polynomial
time.
\end{mytheorem}

\textbf{Proof:} We will show that an algorithm that solves a four,
five and seven link-disjoint path problem solves Problem 2-CESB.
Since finding four, five or seven link-disjoint paths can be solved
by the polynomial time algorithm presented in \cite{seymour},
Problem 2-CESB can be also solved in a polynomial time (we observe
however that the algorithm in \cite{seymour} needs a unrealistic
computation time for determining constants). We focus first on the
case where two of the four special nodes $\{s_1, s_2, t_1, t_2 \}$
are identical, let assume $s_1=s_2=s$. If a solution exists, we can
built another solution that follows one this two compositions:
either four link-disjoint paths to the destinations (no
superposition between the backup paths) or three link-disjoint paths
to the destination and an additional link-disjoint path from one of
the backup path to the destination (when backup shared common
links). In this last composition, the backup paths share a first
partial path and then use different links to get to their respective
destination.

In the second composition, if the backup paths share common links
(but not necessarily continuously) a feasible solution can be
composed by recomposing one of the backup path as follows. If we
denote by $x$ the last node shared by both backup paths before
getting to their respective destination, the partial path between
$s$ and $x$ of the backup path $b_2$ can be modified such that it
uses all the links used by the other backup path $b_1$ until $x$.
This solution is also feasible since there is no overlap with any
primary paths. In order to verify the existence of a solution in
this case where $s_1=s_2$, we can verify the existence of four link
disjoint paths to the destinations by linking each of them twice to
an fictitious node and to run a flow algorithm (first composition).
For the other composition, we verify the existence of five link
disjoint paths (from $s$ to $t_1$, from $s$ to $t_2$, from $s$ to
$x$, from $x$ to $t_1$ and from $x$ to $t_2$) by running an
algorithm that solves the five link-disjoint path problem, testing
it for all $x \in V-\{s_1, s_2, t_1, t_2 \}$. If any solution
exists, the algorithm returns a solution and conversely, any
solution returned by the algorithm solves Problem 2-CESB.

We assume here that the four special nodes $s_1, s_2, t_1$ and $t_2$
are distinct. By a similar analysis, if a solution exists, two
compositions for a solution are relevant: either four link-disjoint
paths from their respective source to their respective destination,
or seven link-disjoint paths (from $s_1$ to $t_1$, from $s_2$ to
$t_2$, from $s_1$ to $x$, from $s_2$ to $x$, from $x$ to $y$, from
$y$ to $t_1$ and from $y$ to $t_2$) where $x$ and $y$ are distinct
regular nodes, as depicted in Figure~\ref{fig:7disjoint}
\begin{figure}
\centering
\includegraphics[width=\smallgraphsize  \textwidth]{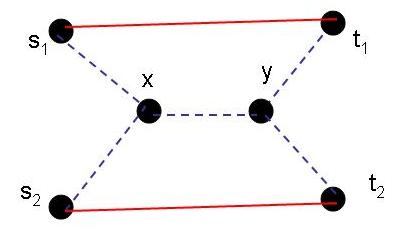}
\caption{The second composition in Problem 2-CESB.}
\label{fig:7disjoint}
\end{figure}
(we observe moreover that we can interchange $s_2$ and $t_2$ since
the graph is undirected). Indeed, if the backup paths in the
original solution have no common links, it corresponds to the first
composition. On the contrary, if some links are shared, we call $x$
and $y$ the first and last nodes shared by $b_2$ with $b_1$ in its
way to $t_2$. We can replace the partial path of the original second
backup path $b_2$ between $x$ and $y$ by the links of $b_1$ in order
to get a feasible solution which corresponds to the second
composition previously described. The first composition can be
solved by an algorithm solving the four link-disjoint path problem,
and the second composition by an algorithm solving the seven link
disjoint path problem.

In all the previous compositions, a solution is returned by the
algorithm if the instance is solvable, and if an algorithm finds a
solution, it corresponds effectively to a solution for Problem
2-CESB. \proofbox

\begin{mytheorem}
\label{thm: Th 3} For a bounded $K$, Problem K-CESB can be resolved
in a polynomial time.
\end{mytheorem}

\textbf{Proof:} We have previously seen the proof for $K=2$. In that
proof, we have seen that an algorithm solving the $l$ link-disjoint
path problem, for $1\leq l\leq 7$, suffices to solve Problem K-CESB
for $K=2$. In this proof, we will generalize it to an algorithm
solving the $l$ link-disjoint path problem, for $1\leq l\leq 5K-3$.
We proceed by strong induction. The property we aim to prove is
formulated as follows: for a bounded $K$, Problem K-CESB can be
solved with the help of an algorithm solving the $l$ link-disjoint
path problem, for $1\leq l\leq 5K-3$. The initialization step is
already verified for $K=1$ and $K=2$.

Let assume that for all $1\leq i\leq K-1$ the property is true for
$i$ and we want to prove that the property is also true for $K$. We
claim that if a solution exists, then a relevant solution can be
found such that it has one of the two following composition: either
a solution with a disconnected backup subgraph (the subgraph
composed of all the backup paths) such that each connected component
corresponds to a case already seen, or a connected backup subgraph
such that all the backup paths are included in it and this subgraph
can be represented by a tree composed at the most of $5K-3$
branches.

We consider a solution to the K-CESB problem and we build the backup
graph. If the graph is connected, we want to show that an algorithm
solving the $l$ link-disjoint path problem, for $1\leq l\leq 5K-3$,
solves our problem. Since all the paths are entirely included in
this connected backup graph, it necessarily includes all the special
nodes $\{ s_1,t_1,..,s_K,t_K \}$. From this connected graph, we can
remove links in order to build a tree that links all these special
nodes such that all the leaves of this tree will be only special
nodes. We consider the number of branches of this tree: in the worst
case, all the special nodes are leaves which are linked to the trunk
through distinct nodes. In the worst case, the number of nodes in
the tree is $2K+(2K-2)=4K-2$, and the number of branches is $4K-3$.
Thus, in order to verify if such a tree exists, we should verify all
the possible combinations for the number of internal nodes (from $1$
to $2K-2$), and all the combinations from these nodes to the special
nodes, such that each special node is linked to exactly one of these
nodes. In other words, we verify the possibility of any tree as
described above by fixing only the internal nodes with a degree
superior to three. In addition, in these verifications, we always work
under the constraint that additional $K$ link-disjoint primary paths
connects their respective source-destination pair. In the worst
case, we will have to find these $K$ primary paths and the $4K-3$
other paths composing the described tree.

If the graph is disconnected, it means that we can find at least two
backup paths such that it is impossible to go from a node of a path
to a node of the other path only through backup links (links used in
this solution by backup paths). In fact, each connected component is
composed of some backup paths, such that each backup path is located
to exactly one connected component. We proceed in a similar way (by
constructing a tree in each connected component). As a consequence,
we can verify if such a solution exists by considering all the
possible partitions of backup paths and all the possible
combinations for each connected component. For each possibility, the
number of link-disjoint paths to consider for a tree of a component
including $j$ backup paths is less than $4j-3$. Thus, the total
number of link-disjoint paths to consider will be also bounded here
by $5K-3$, which verifies the property for $K$. We underline here
that all the considered configurations can be easily verified by the
reader as representing a feasible solution if they exist.

In addition, the total number of combinations $N$ we consider
remains polynomial in $n$. Indeed, we can express an upper bound by:
\begin{eqnarray}
N & \leq & \sum_{m=1}^K \sum_{i_1 \geq .. \geq i_m, s.t. \sum_{l=1}^m i_l =K} \frac{K!}{\prod_{l=1}^m i_l!}.\prod_{i_j \neq 1, j=1}^m \left[\sum_{x=2}^{2 i_j-2} {n \choose x} x^{2 i_j} \right] \nonumber \\
& \leq & K.2^K.n^{2K}.(2K-2)^{2K} \label{eq:UB}\
\end{eqnarray}
We conclude that $N$ is polynomial in $n$. Thus, the algorithm
solving Problem K-CESB verifies these $N$ combinations. If a
solutions exists, the algorithm returns a feasible solution.
Conversely, any solution returns by the algorithm solves Problem
K-CESB. \proofbox

\begin{mytheorem}
\label{thm: Th G2} For a fixed $K$, Problem OffKCP can be solved
within a polynomial number of steps.
\end{mytheorem}

\textbf{Proof:} Problem OffKCP is equivalent by its definition to
Problem K-CESB, which is proven above (Theorem~\ref{thm: Th 3}) to
be solvable within a polynomial number of steps. \proofbox

\begin{mytheorem}
\label{thm: Th 4} For an unfixed $K$, Problem K-CESB is NP-hard.
\end{mytheorem}

\textbf{Proof:} A simple reduction to the general link-disjoint path
problem, which is NP-hard as proven amongst the ``classic'' NP-Hard
problems in \cite{karp} can be done. Considering an instance
$(G,s_1,t_1,..,s_K,t_K)$ of Problem K-CESB with $G=(V,E)$, and the
special nodes $\{(s_1,t_1),..,(s_K,t_K) \}$ in $V$, with $K$ not
necessarily bounded, we construct for the general link-disjoint path
problem the following instance. We add two nodes, $x$ and $y$, and
$2K+1$ links, linking $x$ to all the sources, $y$ to all the
destinations, and $x$ to $y$. This reduction can be done within a
polynomial number of steps. On one hand, an algorithm that would
solve Problem K-CESB would solve the general link-disjoint path
problem by taking the additional links as part of the shared backup
paths. Conversely, from any solution of the general link-disjoint
path problem we can construct a solution to this modified instance
for the problem. \proofbox

\begin{mytheorem}
\label{thm: Th F2} For a general  (not $O(1)$) $K$, Problem KCP is
NP-complete. Moreover, it is NP-hard  to approximate it within a
constant factor.
\end{mytheorem}

\textbf{Proof:} Consider an instance $(G,s_1,t_1,..,s_K,t_K)$ of
Problem KCP, such that $G=(V,E)$ is an undirected uniformly weighted
graph and $\{(s_i,t_i), 1 \leq i \leq K\}$ are the
source-destination pairs of the $K$ connections. We consider Problem
KCP when all the connections are required to be perfectly reliable
(i.e. $MCFP(c_1)=..=MCFP(c_K)=0$). Thus, only the shared backup case
can be applied. For this instance, Problem KCP is equivalent to
Problem K-CESB, which is proven above to be NP-Hard and, moreover,
it is NP-hard to approximate it within a constant factor. \proofbox

\subsection{Problem KCP with a Bandwidth Constraint}

We proceed to establish the intractability of Problem OKCP in the
presence of bandwidth constraints. Even with this bandwidth
constraint, the primary (resp. backup) path of a new connection
request is not allowed to use links already used by a previous
primary or backup (resp. primary) paths. In addition, similarly to
\cite{mira}, the routing does not tolerate traffic splitting.

\begin{definition}
For an instance $I=(G, s_k,t_k,b_k)$, where $G=(V,E)$ is a weighted
undirected graph, we define the weight $b_e$ of link $e$ as composed
of three parameters $(b_1(e), b_2(e)b_3(e))$, such that $b_1(e)$,
$b_2(e)$ and $b_3(e)$ respectively correspond to the bandwidth taken
by the previous primary paths, by the previous backup paths and the
available bandwidth on the link. The capacity $c(e)$ of link $e$ is:
$c(e)=b_1(e)+b_2(e)+b_3(e)$. $(s_K,t_K)$ are two nodes of $V$, $b_k$
is the bandwidth demand of the request, and we call \textit{the
bandwidth online $K$ connection problem} (in short, BW-OKCP) the
problem of finding two link-disjoint paths from $s_K$ to $t_K$
(``primary'' and ``backup'') such that the primary (resp. backup)
path is allowed to use only links with a first and second (resp.
first) bandwidth parameter equal to zero and a third parameter
(resp. the sum of the second and the third parameters) superior to
$b_k$.
\end{definition}

\begin{mytheorem}
\label{thm: Th H} Problem BW-OKCP is NP-complete.
\end{mytheorem}

\textbf{Proof:} We prove that we can reduce Problem BW-OKCP to
Problem 2-DPFL, which is proven above (Theorem~\ref{cor: Co 1}) to
be NP-complete.

We consider an instance $(G,s_K,t_K,b_k)$ of Problem BW-OKCP. We
call $G_1$ the restricted graph of links $e$ of $G$ such that
$b_1(e)=0$. These links are not used by any previous primary paths.
We assume that all the links $e$ of $G_1$ such that $b_2(e)=0$ have
a third parameter $b_3(e)>b_K$. In addition, we assume that amongst
the links $e$ of $G_1$ with $b_2(e)>0$, some links may not have
enough bandwidth $c(e)<b_k$, while others may have. We call $G_2$
the graph resulting from the removal of the links without enough
bandwidth from $G_1$. Thus, Problem BW-OKCP is reduced for this
instance to Problem 2-DPFL, with the instance $(G',s',t',F')$ such
that $G'=G_2$, $s'=s_k$, $t'=t_k$ and the forbidden link set $F'$ is
equal to the set of links in $G_2$ with a strictly positive second
parameter. Indeed, Problem BW-OKCP exactly corresponds in finding
two link-disjoint paths from $s'$ to $t'$ in $G'$ such that only the
backup path is allowed to use links of $F'$. \proofbox

\subsection{The Study of Potential Restrictions for Problem 2CP}
We analyze here the different cases for Problem 2CP. Except common
links between $p_1$ and $p_2$, we consider all the eight
combinations whether we allow superposition between $b_1$ and $b_2$,
$b_1$ and $p_2$, or $p_1$ and $b_2$. We analyze the different
schemes in order to observe if some are superfluous. The following
table sums up the different cases, where only the first three cases
are necessary (Table~\ref{tab:2CPcomp}).
\begin{table}[t]
\begin{center}
\begin{tabular}{|c||c|c|c|}
\hline
Overlapping & ($p_1$,$b_2$) & ($b_1$,$p_2$)  & ($p_2$,$b_2$) \\
\hline
The shared backup case & X &  X & X,V \\ 
The unavoidable first backup case & X &  V & V \\ 
The overlapped connection case & V &  V & X \\ 
The fourth case & V & V & V \\ 
The fifth case & X &  V & X \\ 
The sixth backup case & V &  X & X \\ 
The seventh backup case & V &  X & V \\
\hline
\end{tabular}
\newline\caption{Comparison of the different cases for Problem 2CP.}
\label{tab:2CPcomp}
\end{center}
\end{table}

We detail the features of each case. The first case corresponds to
the shared backup case, when only $b_2$ (or $p_2$ in the symmetric
fifth case) can have superposition with $b_1$, and no one with
$p_1$. We have already seen above that if $p_1$ is fixed, and there
is a feasible solution to this case, then it is the optimal one. The
second case is the unavoidable first backup case, where both paths
of the second connection have a superposition with $b_1$, and none
with $p_1$.

The third case is, like the second case, an interesting one only
when the first case has no feasible solution and then it may be
better than the second case; it has been called the overlapped
connection case.

The next cases (from the fourth to the seventh one) can always be
improved by previous cases and are thus not relevant. In the fourth
case, we take into account both additions in the optimization
expression and get as a goal the minimization of $P_f(p_2)+
P_f(p_1)$. We show that this fourth case cannot outperform the
unavoidable first backup case. If $p_1$ is fixed for example, we get
the same value in the unavoidable first backup case by taking the
shortest path for $p_2=b_2$, and any path for $b_1$ ($b_1$ may have
superposition with $p_2$ and $b_2$). The general case is also
reachable by the unavoidable first backup case by adopting the same
primary path $p_1$.

The fifth case has been showed to be equivalent to the shared backup
case. The sixth and the seventh cases can be improved by the shared
backup case. Since $p_2$ has no superposition with $p_1$ and $b_1$,
only if its links fail then the second connection automatically
fails. The optimization corresponds to the minimization of
$P_f(p_2)$. We can reach the same value by taking $b_2=p_2$ in the
shared backup case: then $b_2$ has no superposition with $p_1$ and
$b_1$, and we minimize the same expression.

\subsection{The Two link-disjoint Path Problem and its Extension}

\begin{definition}
We call {\em the two link path problem} (termed {\em TPP}) the
problem of finding for an instance $I=(G, s_1,s_2,t_1,t_2)$, where
$G$ is an undirected graph and $(s_1,s_2,t_1,t_2)$ are four nodes of
this graph, two link-disjoint paths respectively from $s_1$ to $t_1$
and from $s_2$ to $t_2$.

\end{definition}

We present here the theorem as stated in \cite{frank} that
completely characterizes the solvability of Problem TPP for a
two-link connected graph. The proof can be found in
\cite{thomassen}.

\begin{mytheorem}
\label{thm: Th 5} Let $G$ be a graph such that no cut link separates
both of the two terminal pairs $(s_1,t_1)$ and $(s_2,t_2)$. There is
no two link-disjoint paths between the corresponding terminal pairs
if and only if some links of $G$ can be contracted so that the
resulting graph $G'$ is planar, the four terminals have degree two
while the other nodes are of degree three, and the terminal are
positioned on the outer face in this order: $s_1,s_2,t_1,t_2$.
\end{mytheorem}



}
{}
\end{document}